\documentclass{aa}  

\usepackage{graphicx}
\usepackage{txfonts}
\usepackage[colorlinks=true, citecolor=blue, linkcolor=blue, urlcolor=blue]{hyperref}
\usepackage[dvipsnames]{xcolor}

\begin{document}

   \title{Astrometric modeling of unresolved variable binary systems}
   \subtitle{II.\ Application to Gaia epoch astrometry of nearby pulsating and convective red giants}

   \author{L. Decin\inst{1}\fnmsep\thanks{Corresponding author: leen.decin@kuleuven.be}
   	\and K. Sivkova\inst{2,3}
     \and P. Kervella\inst{3,2}
     \and A. Chiavassa\inst{4}
     \and E. Beguin\inst{4}
        }

   \institute{Institute of Astronomy, KU Leuven, Celestijnenlaan 200D, 3001, Leuven, Belgium
   	\and LIRA, Observatoire de Paris, Université PSL, Sorbonne Université, Université Paris Cité, CY Cergy Paris Université, CNRS, 5 place Jules Janssen, 92195 Meudon, France
    \and French-Chilean Laboratory for Astronomy, IRL 3386, CNRS and U. de Chile, Casilla 36-D, Santiago, Chile
    \and Universit\'e C\^ote d’Azur, Observatoire de la o\^ote d’Azur, CNRS, Laboratoire Lagrange, France
   }

   \date{Received 12 June 2026, Accepted 26 August 2026}

 \abstract
 {The interpretation of high-precision astrometry for intrinsically variable stars remains challenging, particularly for unresolved binary systems containing asymptotic giant branch (AGB) stars. In such systems, large-amplitude pulsations and evolving convective surface structures induce time-dependent photocentre displacements that perturb the observed orbital motion. At the same time, Gaia Data Release 4 and 5 (DR4/DR5) epoch astrometry offers the prospect of deriving accurate orbital solutions and parallaxes for nearby unresolved AGB binaries.}
 {We investigate whether Gaia DR4/DR5 epoch astrometry can reliably recover orbital and astrometric parameters for unresolved AGB binaries in the presence of pulsations and convection-induced photocentre variability.}
 {Building upon the variability-induced mover (VIM) framework, we extend the astrometric model for unresolved evolved binaries by combining Keplerian photocentre motion, pulsation-induced flux-dependent photocentre shifts, and convection-induced photocentre motion. In the forward simulations, convection is represented either by stochastic photocentre displacements drawn from an exponential correlation function or by photocentre time series extracted from three-dimensional radiation-hydrodynamic simulations. We then test a retrieval framework that fits the orbital and VIM signal while incorporating a red-noise covariance matrix to account for correlated astrometric residuals produced by convection.}
 {We show that Gaia epoch astrometry can recover reliable orbital and astrometric parameters for unresolved AGB binaries despite strong pulsation- and convection-induced photocentre variability. Pulsation-induced variability produces a coherent VIM signal that can be modeled jointly with the Keplerian photocentre orbit, while convection-induced photocentre motion behaves primarily as temporally correlated astrometric noise. Retrievals that ignore this correlated component lead to biased proper motions and parallaxes, whereas the inclusion of a physically motivated red-noise covariance model enables accurate recovery of the underlying orbital solution and astrometric parameters for both $\pi^1$~Gru and V~Hya. The retrieval framework remains robust even when the injected convection signal is taken directly from three-dimensional radiation-hydrodynamic simulations rather than from stochastic realizations consistent with the red-noise covariance model adopted in the fit. We also recover the red noise properties.}
 {Our results demonstrate that Gaia DR4/DR5 will provide a powerful new probe of nearby unresolved AGB binaries. The simulations further indicate that correlated-noise treatments, which are currently not part of the Gaia non-single-star processing pipeline, may be essential for accurately characterising evolved variable binaries affected by convection-induced photocentre variability. Accurate parallaxes for these systems also strengthen the potential of long-period variable stars as complementary distance indicators for older stellar populations.}

   \keywords{Astrometry --
                Methods: data analysis --
                binaries: general --
                Stars: variables: general
               }

   \maketitle
   \nolinenumbers

\section{Introduction}~\label{Sec:Intro}

The fourth Gaia data release (DR4) will provide epoch astrometry for the first time, opening a new opportunity for the characterisation of unresolved binary systems through their photocentric motion. The Gaia non-single-star (NSS) processing in DR3 already represents a major advance, providing orbital and other non-single-star solutions for nearly one million sources \citep{Gaia, Gaia2023A&A...674A..34G, Halbwachs_2023}. Nevertheless, this constitutes only a small fraction of the $\sim$1.5 billion sources with a five-parameter astrometric solution in DR3. Given the high intrinsic multiplicity fraction of stars, which is at least $\sim$50\% for Sun-like stars and increases towards higher stellar masses \citep{Moe2017ApJS..230...15M, Fulton2018AJ....156..264F, Offner2023ASPC..534..275O}, a large population of binaries is therefore expected to remain unidentified in DR3 and to be represented by single-star astrometric solutions. This incompleteness is further reinforced by the limited time baseline and number of observations available in DR3, as well as by the stringent selection criteria adopted to ensure the reliability of the published NSS solutions \citep{Gaia2023A&A...674A..34G, Halbwachs_2023}. For such unresolved systems, unmodeled orbital motion can affect the inferred astrometric parameters, most notably the proper motion. The parallax is generally more robust owing to its characteristic annual signature, although some bias may occur when the orbital or other photocentric motion is partially degenerate with the parallactic motion. For intrinsically variable stars, the situation becomes even more complex because photometric variability induces additional time-dependent photocentre displacements.

\citet[][hereafter Paper~I]{PaperI} developed a simulation framework to investigate the impact of unresolved orbital motion and variability-induced motion (VIM) on Gaia epoch astrometry, with particular emphasis on binary Cepheids. VIM systems are unresolved binaries containing one photometrically variable component, typically a pulsating star \citep{Wielen_1996}. As the flux contribution of the variable component changes over time, the photocentre of the system shifts correspondingly. Because Gaia combines high-precision astrometry with simultaneous photometric monitoring, Gaia epoch astrometry now makes it possible to disentangle orbital motion from pulsation-induced photocentre shifts.

Cepheids provide a natural first application of the VIM framework because they are both intrinsically variable and frequent members of binary or multiple systems \citep{Evans2015AJ....150...13E,Kervella2019A&A...623A.116K}. Their importance further stems from their central role in the cosmic distance ladder through the period--luminosity relation \citep{Riess2021ApJ...908L...6R, Breuval2024ApJ...973...30B}. In Paper~I, it was shown that unresolved orbital motion may bias Gaia astrometric solutions for binary Cepheids if the systems are modeled as single stars.

A similar, and potentially even more challenging, situation arises for asymptotic giant branch (AGB) stars. 
Most stars with initial masses above $\sim$0.8\,M$_\odot$ likely host at least one stellar or planetary companion~\citep{Moe2017ApJS..230...15M,Fulton2018AJ....156..264F,Decin2020Sci...369.1497D}, and $\sim$95\% are expected to evolve through the AGB phase. 
Long-period variable AGB stars also follow period--luminosity relations \citep{Whitelock2008MNRAS.386..313W,Yuan2017AJ....154..149Y,Huang2020ApJ...889....5H} and have recently been proposed as complementary distance indicators for old stellar populations inaccessible to classical Cepheids \citep{Huang2018ApJ...857...67H,Trabucchi2021A&A...656A..66T, Huang2024ApJ...963...83H,H0DN2026A&A...708A.166H}. In contrast to Cepheids, however, AGB stars exhibit not only large-amplitude pulsations but also strong surface convection. Three-dimensional radiation-hydrodynamic simulations predict that evolving convective surface structures can induce stochastic photocentre displacements at the sub-milliarcsecond level, potentially perturbing Gaia astrometry even for single stars \citep{Chiavassa2018A&A...617L...1C,Beguin_2024}. 
Whether such convection-induced photocentre motion can already be identified in Gaia catalogue-level astrometric diagnostics remains debated. While \citet{Chiavassa2022A&A...661L...1C} reported evidence for convection-induced astrometric jitter in red supergiants based on Gaia EDR3 parallax uncertainties, \citet{Kochanek2023MNRAS.520.3510K} questioned this interpretation, finding that the EDR3 astrometric excess noise of red supergiants is approximately an order of magnitude larger than the expected convection-induced signal and is not distinguishable from that of comparison stars with similar observational properties.
Large-scale convective structures have also been directly imaged or inferred
observationally on nearby evolved stars through optical/infrared
interferometry and ALMA monitoring
\citep{Paladini2018Natur.553..310P,Chiavassa2020A&A...640A..23C, Vlemmings2024Natur.633..323V}.  In unresolved binary systems, these stochastic photocentre shifts coexist with coherent variability-induced motion and Keplerian orbital motion, complicating the interpretation of the astrometric signal.

At present, only a very limited number of nearby AGB binaries possess sufficiently well-constrained orbital architectures and stellar properties to enable realistic forward simulations of Gaia epoch astrometry. Rather than performing a broad population-synthesis study based on uncertain assumptions regarding pulsation, convection, and companion properties, we  adopt a proof-of-concept approach using the two best-characterised compact AGB binary systems currently known: $\pi^1$~Gru and V~Hya. These systems provide particularly valuable benchmark cases because independent observational constraints already exist on their orbital configuration and variability properties from ALMA, interferometric, spectroscopic, photometric, and Hipparcos--Gaia analyses \citep[e.g.][]{Paladini2018Natur.553..310P, Planquart2024A&A...682A.143P,Esseldeurs2026NatAs..10..124E,Montarges2025A&A...699A..22M}.

In this work, we extend the VIM framework introduced in Paper~I to unresolved binaries containing evolved giant stars. Pulsation-induced photometric variability is incorporated through the VIMO formalism, while stochastic convection-induced photocentre variability is treated separately as a correlated astrometric noise component. We therefore distinguish between classical variability-induced motion (VIM), driven by coherent pulsation-induced flux variations in unresolved binaries, and convection-induced photocentre motion (CIM), which may also affect single stars and is modeled statistically through a red-noise covariance formalism. Using simulated Gaia DR4- and DR5-like epoch astrometry for $\pi^1$~Gru and V~Hya, we investigate under which conditions reliable orbital solutions and parallaxes can be recovered in the presence of strong intrinsic variability. Although these nearby benchmark systems are not intended to represent the full population of AGB stars, they probe the types of astrometric biases that unresolved variability and convection may introduce in Gaia parallaxes of long-period variables.

This paper is organized as follows. Sect.~\ref{Sec:Method} summarizes the astrometric formalism and the implementation of pulsations and convection-induced photocentre variability. Sect.~\ref{Sec:benchmark} presents the simulated Gaia epoch astrometry and the retrieval analysis for $\pi^1$~Gru and V~Hya. In Sect.~\ref{Sec:Discussion} we discuss the implications for Gaia astrometric solutions and we present our conclusions in Sect.~\ref{Sec:conclusions}.

\section{Modelling Gaia epoch astrometry of variable AGB binaries}
\label{Sec:Method}

Gaia measures stellar positions in a one-dimensional reference
frame defined by the satellite scanning law. Rather than
recording direct two-dimensional positions on the sky, each
Gaia transit provides a highly precise along-scan (AL)
measurement corresponding to the projection of the source
position onto the instantaneous scanning direction
\citep{Lindegren_2012}. The Gaia epoch astrometry of a source
therefore consists of a time series of along-scan measurements
obtained at different scan angles and parallax factors over the
mission lifetime.

A detailed description of the Gaia along-scan formalism and
the corresponding forward-simulation pipeline is presented in
Paper~I. Here, we summarize only the elements required for the
modelling of unresolved variable AGB binaries, including
Keplerian photocentre motion, pulsation-induced variability,
and convection-induced photocentre displacements.

To construct realistic Gaia-like epoch astrometry, we adopt
the Gaia observing cadence predicted by the Gaia Observation
Schedule Tool (GOST)\footnote{\url{https://gaia.esac.esa.int/gost/}},
which provides the Gaia observing epochs, scan angles, and
parallax factors corresponding to a given sky position.
Synthetic along-scan astrometric measurements are then
generated by projecting the simulated photocentre motion onto
the Gaia scanning direction at each observing epoch.

Following Paper~I, Gaussian Gaia-like measurement
uncertainties are added using the magnitude-dependent
per-CCD uncertainty prescription described by
\citet{ElBadry2024OJAp....7E.100E}, evaluated at the Gaia
$G$-band magnitude of the unresolved system.

\subsection{Single-star astrometric signal simulation}

For a single star, the Gaia along-scan coordinate can be written as
\begin{equation}
	w_{\rm SS}
	=
	(\Delta \alpha^\star + \mu_{\alpha^\star} t)\sin\theta
	+
	(\Delta\delta + \mu_\delta t)\cos\theta
	+
	\Pi \varpi,
\end{equation}
where $\Delta \alpha^\star$ and $\Delta\delta$ denote the
offsets relative to the reference position
$(\alpha_0,\delta_0)$,
$\mu_{\alpha^\star}$ and $\mu_\delta$ are the proper-motion
components,
$\varpi$ is the parallax, and
$\Pi$ the along-scan parallax factor.
The scan angle $\theta$ defines the orientation of the Gaia
along-scan direction at the observing epoch, while
$t=t_{\rm obs}-t_{\rm ref}$ is measured relative to the Gaia
reference epoch. 
The colour of the source, and potentially its time variability, has an effect on the five-parameter astrometric solutions of Gaia and thus the parallax. This can be corrected through the introduction of a sixth fitted parameter, the effective wavenumber $\nu_\mathrm{eff}$, as detailed in \citet{Lindegren2021A&A...649A...4L}. In the present study, we do not consider this additional parameter, whose effect on the Gaia data is expected to be small compared to other stochastic perturbations. However, it will eventually be possible to take it into account using the $G_\mathrm{BP}-G_\mathrm{RP}$ epoch photometry that will be provided in the Gaia DR4.

\subsection{Binary astrometric signal simulation}

\subsubsection{Keplerian photocentre motion}

For an unresolved binary system, the along-scan astrometric
signal additionally contains the projected motion of the system
photocentre induced by the orbital motion of the binary. The
corresponding along-scan coordinate becomes
\begin{equation}
	\begin{aligned}
		\label{w_bs}
		w_{\mathrm{bs}}
		&=
		(\Delta \alpha^\star
		+
		\mu_{\alpha^\star} t
		+
		\alpha^\star_{\mathrm{orb}}(t))
		\sin \theta
		\\
		&+
		(\Delta \delta
		+
		\mu_{\delta} t
		+
		\delta_{\mathrm{orb}}(t))
		\cos \theta
		\\
		&+
		\Pi \varpi
		\\
		&=
		w_{\mathrm{SS}}
		+
		\alpha^\star_{\mathrm{orb}}(t)\sin\theta
		+
		\delta_{\mathrm{orb}}(t)\cos\theta,
	\end{aligned}
\end{equation}
where $w_{\mathrm{SS}}$ now represents the barycentric motion, and 
$\alpha^\star_{\mathrm{orb}}(t)$ and
$\delta_{\mathrm{orb}}(t)$ describe the photocentre motion on
the plane of the sky induced by the unresolved binary orbit and,
in the case of variable systems, by variability-induced
photocentre displacements (see Sect.~\ref{Sec:VIM}).

For an unresolved binary, the semi-major axis of the
photocentre orbit is related to the relative Keplerian orbit
through
\begin{equation}
	a_{\rm ph}
	=
	a_{\rm rel}\,|B-\beta|,
	\label{Eq:a_ph}
\end{equation}
where $a_{\rm rel}$ is the semi-major axis of the relative
orbit, $B = M_2/(M_1 + M_2)$ is the fractional mass contribution of the companion, with
$M_1$ and $M_2$ the masses of the primary and secondary,
respectively, and $\beta =  F_2/(F_1 + F_2) = F_2/F$ 
is the fractional flux contribution of the companion, where
$F_1$ and $F_2$ denote the fluxes of the primary and secondary,
respectively, and $F$ is the total system flux.

\subsubsection{Variability-induced motion} \label{Sec:VIM}

In variability-induced movers (VIMs; \citealt{Wielen_1996}),
one component of the unresolved binary system is
photometrically variable due to pulsations. The resulting time-dependent flux
variations modify the photocentre position and induce an
additional astrometric signal coupled to the photometric
variability. In this case, the photocentre position can be expressed as
\begin{equation}
	x_{\rm ph}(t)
	=
	\frac{
		F_1(t)x_1(t)
		+
		F_2x_2(t)
	}
	{
		F_1(t)+F_2
	},
\label{Eq:x_ph}
\end{equation}
where $x_i(t)$ denotes the position of component $i$
projected onto either $\alpha^\star$ or $\delta$,
$F_1(t)$ is the time-dependent flux of the variable
primary component, and $F_2$ the (assumed constant) flux of the companion over the observing baseline.
The total system flux therefore becomes $F(t)=F_1(t)+F_2$.
As a consequence, the photocentre motion no longer follows
a purely Keplerian orbit. Instead, the photocentre
semi-major axis itself becomes time dependent,
$a_{\rm ph}(t)$.

Paper~I discussed the hierarchy of simplified VIM models
introduced by \citet{Wielen_1996}, namely the fixed (VIMF),
linear (VIML), and acceleration (VIMA) approximations.
In the present work, we focus primarily on the full
Keplerian variability-induced mover formalism (VIMO),
which is required to model the large photocentre
excursions expected in nearby AGB binaries.

\subsubsection{VIMO formalism} \label{Sec:VIMO}

The simplified VIMF, VIML, and VIMA descriptions provide low-order approximations to the photocentre motion when the orbital signal is only partially resolved. For the nearby AGB binaries considered here, however, the Gaia observing window samples a substantial fraction of the orbit, while the pulsation-induced photocentre displacement can have a major contribution to the astrometric signal. We therefore use the full orbital variability-induced mover model (VIMO; \citealt{Wielen_1996}), in which the variability-induced motion is coupled directly to the Keplerian relative orbit.

Following \citet{Wielen_1996}, the photocentre position (Eq.~\ref{Eq:x_ph})  can be written as
\begin{eqnarray}
	x_{\rm ph}(t) & = & x_{\rm CoM} + \dot{x}_{\rm CoM}\,t - (B-\beta(t))\, a_{\rm rel} s_x(t) \\
	& \equiv & x_{\rm CoM} + \dot{x}_{\rm CoM}\,t - (B - \overline{\beta})\, a_{\rm rel} s_x(t) \nonumber \\
	& & + \overline{\beta} a_{\rm rel} f(t) s_x(t)
\end{eqnarray}
where $x$ is the position in $\alpha^\star$ or $\delta$, $x_{\rm CoM}$ is the position of the centre-of-mass of the binary at $t=t_{\rm ref}$ with $\dot{x}_{\rm CoM}$  its proper motion, $s_x(t)$ describes the normalized orbital motion of the binary for the set of orbital elements, and we define 
\begin{equation}
	\overline{\beta} = \frac{F_2}{\overline{F}} = \frac{F_2}{\overline{F_1(t)} + F_2}
	\label{Eq:beta_bar}
\end{equation}
as the flux ratio $\beta$ for a given reference flux $\overline{F}$ (here taken as the average of $F(t)$). Hence, the orbital contribution can be written as
\begin{eqnarray}
	x_{\rm ph}^{\rm orb}(t) & = & \left( -(B-\overline{\beta}) + \overline{\beta} f(t) \right)\, a_{\rm rel} s_x(t) \nonumber \\
	& \equiv & a_{\rm ph} (t) s_x^{\rm ph} (t) \\
	& \equiv & \overline{\alpha}  s_x^{\rm ph}(t)  \left(1 - A_f f(t) \right) \label{Eq:x_ph_VIMO}
\end{eqnarray}
where we define $s_x^{\rm ph}(t)$ as the normalized orbital motion of the photocentre, and $\overline{\alpha}$ the semi-major axis of the photocentre orbit for a reference flux $\overline{F}$, with
\begin{eqnarray}
	\overline{\alpha} & = & |B - \overline{\beta}|\,a_{\rm rel} \label{Eq:alpha_bar}\\
	A_f & = & -{\rm sgn}(B-\overline{\beta})\, \frac{ \overline{\beta} a_{\rm rel}}{\overline{\alpha}} \label{Eq:Af}\\
	f(t) & = & \frac{\overline{F}}{F(t)} - 1 \,.
\end{eqnarray}

\subsection{Pulsation-induced photometric variability} \label{Sec:pulsations}

For nearby AGB stars, the pulsation-induced photocentre displacement can reach amplitudes corresponding to a substantial fraction of the stellar radius and, in some systems, up to order unity relative to the Keplerian photocentre orbit itself. Accurately characterising the pulsation signal therefore becomes essential for disentangling the orbital motion from the variability-induced astrometric contribution.

In principle, Gaia epoch photometry could provide a direct constraint on the pulsation cycle simultaneously with the epoch astrometry. However, the Gaia DR4/DR5 epoch photometry is not yet publicly available, while the currently accessible Gaia DR3 photometric time series remain too sparsely sampled to robustly reconstruct the pulsation cycle of long-period AGB variables. We therefore adopt, in this proof-of-concept study, a parametric description of the pulsation variability.

For each system, we use the Gaia archive values of $G_{\rm min}$ and $G_{\rm max}$ to constrain the variability amplitude of the primary AGB star in the Gaia $G$ band. The photometric variability is modeled using a simplified asymmetric sine prescription characterised by the extrema $G_{\rm min}$ and $G_{\rm max}$, the pulsation period $P_{\rm puls}$, the reference epoch $T_{0,\rm puls}$, and the asymmetry (tilt) parameter $\Gamma$ \citep{Jorissen2004agbs.book..461J,Planquart2024A&A...682A.143P}. The pulsation periods and reference epochs are adopted from dedicated photometric monitoring studies available in the literature. This parametrized light-curve model is then used to construct the normalized variability function $f(t)$ entering the VIMO formalism.

\subsection{Convection-induced photocentre variability}\label{Sec:convection}

In addition to coherent pulsations, AGB stars exhibit large-scale
surface convection that can induce stochastic photocentre
displacements \citep{Paladini2018Natur.553..310P, Chiavassa2018A&A...617L...1C, Vlemmings2024Natur.633..323V, Beguin_2024}. Such photocentre variability  is expected to affect Gaia astrometry even for single stars.

To include this effect in the forward simulations, we considered
two different prescriptions for the convection-induced photocentre
variability. The first approach uses a stochastic
Ornstein--Uhlenbeck (OU) process \citep{UhlenbeckOrnstein1930}, which provides a simple
phenomenological description of surface-convection variability
with finite memory. In contrast to white noise, successive
photocentre displacements remain temporally correlated over
timescales of order $\tau_{\rm conv}$. Conversely, unlike an
unconstrained random walk, the process is mean reverting and
therefore does not produce unbounded photocentre drifts over
long timescales. The OU process therefore provides a natural
first-order approximation for the evolution of large-scale
convective structures on the surface of AGB stars.

The temporal correlation of the stochastic photocentre
displacements is then described by the exponential correlation
function
\begin{equation}
	K(\Delta t)
	=
	\sigma_{\rm conv}^2
	\exp
	\left(
	-\frac{|\Delta t|}{\tau_{\rm conv}}
	\right),
	\label{Eq:OU}
\end{equation}
where $\Delta t$ is the time lag between two epochs.
The correlation timescale $\tau_{\rm conv}$ reflects the typical
lifetime of the dominant convective cells, while
$\sigma_{\rm conv}$ characterises the associated photocentre
excursions generated by the evolving surface-brightness
inhomogeneities.
Hydrodynamical simulations and interferometric monitoring of
nearby AGB stars suggest that the evolution timescale of the
dominant convective cells is of the order of several weeks
\citep{Vlemmings2024Natur.633..323V,Beguin_2024}. We therefore
adopt correlation timescales of this order in our simulations.
The characteristic photocentre displacement amplitude
$\sigma_{\rm conv}$ is estimated using Eq.~8 of
\citet{Beguin_2024}, which relates the expected photocentre
excursions to pulsation period, after correlating it with the stellar radius and surface-brightness
inhomogeneities predicted by three-dimensional
radiation-hydrodynamic simulations.

In addition to the phenomenological OU prescription, we also
performed forward simulations directly using the photocentre
variability predicted by the three-dimensional
radiation-hydrodynamic simulation
\texttt{st29gm04n001} from \citet{Chiavassa2018A&A...617L...1C} and  
\citet{Beguin_2024}\footnote{The three-dimensional radiation-hydrodynamic model
	\texttt{st29gm04n001} from \citet{Chiavassa2018A&A...617L...1C}
	corresponds to a solar-metallicity AGB star with
	$M_\star\!=\!1\,M_\odot$,
	$L_\star\!=\!4982\,L_\odot$,
	$R_\star\!=\!1.37\,{\rm au}$,
	and $T_{\rm eff}\!=\!2827\,{\rm K}$.
	The authors furthermore generated synthetic Gaia observables
	using Gaia-specific filter passbands and instrumental properties
	adapted to the photometric and astrometric response of the
	space mission.}.  In this case, the time-dependent photocentre
displacements predicted by the hydrodynamical simulation were
interpolated onto the Gaia observing epochs and added directly
to the synthetic astrometric signal. This approach provides a
more physically realistic description of the convection-induced
photocentre variability and allows us to test whether the
red-noise retrieval framework (see Sect.~\ref{sec:rednoise}) remains robust when the injected
variability no longer follows the simplified OU-process
assumption.

To approximate the surface-convection effects, we generated independent realizations of right ascension and declination photocentre offsets at the Gaia epochs before projecting them onto the Gaia along-scan direction. We then added the resulting convection-induced photocentre motion to the system VIM simulation.

\subsection{Retrieval methodology}

\subsubsection{Simplified astrometric retrievals}

We first consider the same hierarchy of simplified astrometric
models as introduced in Paper~I. These include the single-star
(SS) model and the fixed, linear, and acceleration
variability-induced mover models, denoted VIMF, VIML, and
VIMA, respectively \citep{Wielen_1996} and include only white-noise in the covariance matrix. 
These models provide low-order descriptions of the photocentre motion and are useful
to quantify the biases that arise when orbital curvature and
variability-induced motion are not fully modeled.
The explicit along-scan expressions and fitting procedure for
the SS, VIMF, VIML, and VIMA models are given in Paper~I.
Here, we use them only as reference retrievals against which the
full Keplerian VIMO model can be compared.

\subsubsection{Keplerian VIMO retrieval}\label{Sec:VIMO_retrieval}

For the full VIMO retrieval, we extend
the \texttt{kepmodel} framework developed by
\citet{kepmodel}. The model follows the VIMO formalism
introduced in Sect.~\ref{Sec:VIMO} and fits the astrometric signal as a
Keplerian photocentre orbit whose amplitude is modulated by
the photometric variability of the primary star.
In the fitting procedure, the observed VIMO signal (Eq.~\ref{Eq:x_ph_VIMO}) is
approximated as
\begin{equation}
	\overline{\alpha}
	s_x^{\rm ph}(t)
	\left(
	1 - A_f f(t)
	\right)
	\simeq
	\hat{\alpha}
	s_x^{\rm ph}(t)
	\left(
	1 - \hat{A} \tilde{f}(t)
	\right),
\end{equation}
where $\overline{\alpha}$ is the true photocentre semi-major axis at reference flux $\overline{F}$,
$\hat{\alpha}$ its estimator,
$s_x^{\rm ph}(t)$ the normalized Keplerian photocentre orbit,
$f(t)$ the photometric modulation term,
and $A_f$ the true variability-induced modulation amplitude.

If both the companion flux and the intrinsic variability model are
known exactly, then
$\tilde{f}(t)=f(t)$ and $\hat{A}$ directly estimates $A_f$.
In the more general case where $F_2$ is unknown, the code assumes a
faint companion with constant flux
$\tilde{F}_2=\delta_2 F_2$,
with $\delta_2 \ll 1$, so that
\begin{equation}
	\tilde{f}(t)
	=
	\frac{
		\overline{F_1(t)}+ \tilde{F}_2
	}{
		F_1(t) + \tilde{F}_2
	}
	- 1\,.
\end{equation}

In practice, the fitting procedure introduces an additional linear
predictor proportional to
$\tilde{f}(t)\,w_{\rm orb}(t)$,
whose coefficient corresponds to $\hat{A}$ and where
$
w_{\rm orb}(t)
=
\alpha^\star_{\rm orb}(t)\sin\theta
+
\delta_{\rm orb}(t)\cos\theta$. 
The nonlinear orbital parameters are optimized simultaneously with the
linear astrometric coefficients through nonlinear least-squares
minimization. The parameter covariance matrix is estimated from the curvature of the
likelihood evaluated at the best-fit solution. In the presence of
correlated noise (see Sect.~\ref{sec:rednoise}), this likelihood uses the full covariance matrix
defined in Eq.~\eqref{Eq:C_white_red}.

Astrometric Keplerian solutions possess an intrinsic branch degeneracy
corresponding to a simultaneous sign inversion of the orbital
semi-major axis and a rotation of $180^\circ$ in both $\omega$ and
$\Omega$. To ensure a unique representation of the solution, we adopted the
convention $\hat{A} \geq 0$. For non-circular solutions, a negative fitted amplitude was mapped to the
equivalent positive-amplitude branch by applying
$\hat{A} \rightarrow -\hat{A}$ together with
$\omega \rightarrow \omega + 180^\circ$ and
$\Omega \rightarrow \Omega + 180^\circ$.
For nearly circular orbits, where $\omega$ is poorly defined, we only
enforced the convention $\hat{A} \geq 0$ and did not interpret the resulting
argument of periastron.

\subsubsection{Red-noise covariance treatment}
\label{sec:rednoise}

The convection-induced photocentre variability introduced in
Sect.~\ref{Sec:convection} produces temporally correlated
astrometric residuals that cannot be represented adequately by
independent white-noise measurements. To account for this effect
in the retrieval, we incorporated the stochastic photocentre
variability through a Gaussian-process covariance matrix.

The covariance matrix of the correlated astrometric component is
constructed from the Ornstein--Uhlenbeck correlation function
defined in Eq.~\eqref{Eq:OU},
\begin{equation}
	[C_{\rm red}]_{ij}
	=
	K(|t_i-t_j|),
\end{equation}
where $t_i$ and $t_j$ denote two Gaia observing epochs.
Separate covariance matrices are constructed for right ascension
and declination before projection onto the Gaia along-scan
direction.

The total covariance matrix is written as
\begin{equation}
	C
	=
	C_{\mathrm{Gaia}}
	+
	C_{\mathrm{jit}}
	+
	C_{\mathrm{red}},
	\label{Eq:C_white_red}
\end{equation}
where $C_{\mathrm{Gaia}}$ contains the formal Gaia along-scan measurement uncertainties -- here based on the per-CCD uncertainties as formulated in \citet{ElBadry2024OJAp....7E.100E}, $C_{\mathrm{red}}$ the correlated convection-induced
covariance matrix, and $C_{\mathrm{jit}}$ an additional
white-noise jitter term intended to absorb unresolved
short-timescale photocentre variability or calibration
imperfections. The jitter term was initialized to 0.050\,mas, corresponding
approximately to the median formal DR3 per-CCD uncertainty
at $G=12$ \citep{Lindegren2021A&A...649A...2L}. When enabled
in the fit, the jitter amplitude was optimized simultaneously
with the astrometric model parameters.

In the presence of correlated noise, the relevant goodness-of-fit
statistic becomes the generalized chi-square
\begin{equation}
	\chi^2_{\rm gen}
	=
	\mathbf{r}^{\rm T}\mathbf{C}^{-1}\mathbf{r},
\end{equation}
where $\mathbf r$ is the residual vector between the observed and
modeled along-scan astrometry. Using the Cholesky decomposition
\begin{equation}
	\mathbf C = \mathbf L \mathbf L^{\rm T},
\end{equation}
the generalized chi-square can be evaluated through the
whitened residuals
\begin{equation}
	\mathbf u = \mathbf L^{-1}\mathbf r,
\end{equation}
such that
\begin{equation}
	\chi^2_{\rm gen}
	=
	\mathbf u^{\rm T}\mathbf u.
\end{equation}
The corresponding generalized reduced chi-square is
\begin{equation}
	\chi^2_{{\rm red,gen}}
	=
	\frac{\chi^2_{\rm gen}}{\nu},
\end{equation}
with $\nu$ the number of degrees of freedom.

For covariance models that include free red-noise
hyperparameters, model comparison was performed using the full
Gaussian likelihood $\mathcal L$ rather than
$\chi^2_{\rm gen}$ alone. The minimized objective function is
\begin{equation}
	{\rm obj}
	=
	\chi^2_{\rm gen}
	+
	\ln \det(\mathbf C),
	\label{Eq:obj}
\end{equation}
which corresponds to $-2\ln\mathcal L$ up to an additive
constant. The determinant term penalizes covariance models with artificially inflated noise amplitudes and discourages the correlated-noise component from absorbing deterministic orbital structure.

When the red-noise hyperparameters were left completely
unconstrained, the covariance optimization occasionally
converged toward very long correlation timescales comparable
to, or even exceeding, the Gaia observing baseline. This
behaviour occurred primarily when fitting overly simplified
astrometric models, such as a single-star solution, to Gaia
epoch astrometry still containing an unresolved Keplerian
orbital component. In that regime, the stochastic covariance
term can partially absorb the low-frequency orbital signal and
therefore becomes strongly degenerate with both the low-order
astrometric parameters and the unmodeled orbital motion.
To avoid such nonphysical solutions, we restricted the
red-noise hyperparameters to physically motivated ranges
representative of surface-convection variability in AGB stars,
namely
\begin{equation}
	20 \le \tau_{\rm conv} \le 300~{\rm d}
\end{equation}
and
\begin{equation}
	0.01 \le \sigma_{\rm conv} \le 2~{\rm mas}.
\end{equation}

For the correlated (red) noise SS and VIMO fits, parameter uncertainties were
estimated directly from the curvature of the full likelihood
using the covariance matrix $\mathbf C$. The reported
uncertainties therefore already account for the adopted
correlated-noise model.

Values of $\chi^2_{{\rm red,gen}}$ close to unity do not
necessarily imply that the physical model provides an adequate
description of the data, since correlated-noise models may
partially absorb unmodeled deterministic structure such as
orbital motion. The interpretation of
$\chi^2_{{\rm red,gen}}$ therefore requires verification that
the retrieved covariance hyperparameters remain physically
plausible and do not converge toward imposed parameter
boundaries.

\section{Benchmark systems: validating the combined VIM and CIM framework on AGB binaries}\label{Sec:benchmark}

The retrieval framework described in Sect.~2 is applied to the
only two currently well-characterised AGB binary systems,
$\pi^1$~Gru and V~Hya, for which sufficiently complete and
independent orbital constraints exist to permit a meaningful
validation of the VIMO retrievals. Both systems have
orbital separations smaller than $\sim$15~au, placing them in a
regime where unresolved photocentre motion becomes particularly
relevant for Gaia astrometry. Their orbital periods are also
well suited to the Gaia mission, since the nominal mission
duration, and especially the expected temporal coverage of
Gaia DR5, samples a substantial fraction of the orbit.

Long-period variability on the AGB is commonly divided into
Mira and semi-regular (SR) variability classes
\citep{Wood2000PASA...17...18W}. Mira stars exhibit large,
highly coherent pulsations, whereas SR variables typically show
lower-amplitude and less regular or multi-periodic variability.
These different variability regimes are expected to produce
markedly different VIM signatures
in Gaia astrometry. The benchmark systems studied here
 probe two complementary regimes: the SR system
$\pi^1$~Gru, in which convection-induced photocentre
variability dominates over the pulsation-induced VIM signal,
and the Mira-type system V~Hya, where pulsation-induced
VIM motion becomes a major component of the
astrometric signal.

\paragraph{$\pi^1$~Gru:} $\pi^1$ Gruis is a triple system containing the main AGB star,
a semi-regular variable of type SRb, a distant G0V-type
companion B separated by 2.8 arcsec from the primary
\citep[Gaia DR3 6518817665842868352;][]{1953MNRAS.113..510F,Kervella_2022},
and a closer companion C.
Here, we considered the $\pi^1$~Gru A--C system, which has an
orbital separation of $\sim$6.6~au
\citep{Esseldeurs2026NatAs..10..124E,Montarges2025A&A...699A..22M}.
Using multi-epoch ALMA observations,
\citet{Esseldeurs2026NatAs..10..124E} spatially resolved the
circum-companion accretion disk associated with companion C and
derived a detailed orbital and astrometric solution combining
ALMA astrometry with Hipparcos and Gaia DR3 measurements of
the primary star. They also derived the barycentric proper
motion of the system, which we adopted in our forward
simulations.

The Gaia $G$ band magnitude of $\pi^1$~Gru~A varies between 3.253 and 3.727 \footnote{\label{gaia_archive}\url{https://gea.esac.esa.int/archive/}}. Using $V$-band data, \citet{Watson2006SASS...25...47W} derived a pulsation period of approximately 195.5~d. \citet{Esseldeurs2026NatAs..10..124E} interpreted this period as the radial first-overtone mode. We  modeled the pulsation variability using the simplified sinusoidal approach described in Sect.~\ref{Sec:pulsations}. 

Companion C remains currently undetected. Assuming an F8V main-sequence companion, as suggested by \citet{Esseldeurs2026NatAs..10..124E}, we estimated a companion magnitude of $V_{\rm comp}=10.2$ for the measured system parallax. This value was adopted in the forward simulations to construct the synthetic photocentre variability signal. However, during the retrieval stage, the companion flux was assumed to be unknown and was therefore not fixed in the VIMO  fitting procedure (see Sect.~\ref{Sec:VIMO_retrieval}).

\paragraph{V~Hya:}
V~Hydrae is a Mira-type AGB star showing two dominant
photometric timescales, namely a Mira-like pulsation period of
$\sim$530~d and a long secondary period of $\sim$17.45~yr
\citep{Planquart2024A&A...682A.143P}, for which
\citet{Planquart2024A&A...682A.143P} derived a Keplerian
orbital solution using twelve years of HERMES radial-velocity
monitoring together with archival visual photometry and
Hipparcos--Gaia astrometric acceleration measurements to
disentangle the pulsational and orbital signals
(see Table~\ref{tab:input_retrieved_params}).
Unlike $\pi^1$~Gru, the companion of V~Hya has not yet been
directly detected spatially, and its presence is currently
inferred indirectly from the combined radial-velocity,
photometric, ultraviolet, and astrometric signatures.
Since no independent barycentric proper motion is currently
available for V~Hya, we applied the DR3 proper-motion correction
described in Paper~I, resulting in a correction of
$\Delta \mu_{\alpha^\star}$\,=\,$-1.693$~mas\,a$^{-1}$ and
$\Delta \mu_{\delta}$\,=\,4.691~mas\,a$^{-1}$, yielding
$\mu_{\alpha^\star}$\,=\,$-4.497$~mas\,a$^{-1}$ and
$\mu_{\delta}$\,=\,$-2.695$~mas\,a$^{-1}$.

The Gaia $G$-band magnitude of V~Hya varies between approximately 5.08 and 7.0~$^{\ref{gaia_archive}}$. We modeled the Mira pulsation variability using the asymmetric sine approach described in Sect.~\ref{Sec:pulsations}, adopting the 530~d pulsation period \citep{Planquart2024A&A...682A.143P}. 

Based on the UV excess of V~Hya, \citet{Planquart2024A&A...682A.143P} argued for a hot main-sequence companion, consistent with an A-type star of temperature $\simeq$\,9\,500~K. For the measured system parallax, this corresponds to an estimated Gaia magnitude of $G_{\rm comp}\simeq8.8$.

\medskip

Due to the pulsation-induced VIM effect, the photocentre
semi-major axis (Eq.~\eqref{Eq:a_ph}) varies over the pulsation
cycle. The maximum displacement of the photocentre between the
minimum and maximum of $F(t)$ is given by
\begin{equation}
	d_{\rm var} \equiv \Delta a_{\rm ph}
	=
	a_{\rm rel} F_2
	\left(
	\frac{1}{F_{\rm min}}
	-
	\frac{1}{F_{\rm max}}
	\right)\,.
	\label{Eq:d_var}
\end{equation}
The resulting total displacement $d_{\rm var}$
is $\sim$0.03~mas for $\pi^1$~Gru and $\sim$5.7~mas for V~Hya,
demonstrating that the variability-induced motion is
substantially larger in the V~Hya system.

For both AGB stars, we followed the OU-process prescription
described in Sect.~\ref{Sec:convection} to approximate the
convection-induced photocentre variability. The characteristic
correlation timescale was fixed to $\tau_{\rm conv}=30$~d,
similar to the timescale inferred for R~Dor by
\citet{Vlemmings2024Natur.633..323V}. Assuming a stellar radius
of $R_\star\!\sim\!1.65$~au for $\pi^1$~Gru
\citep{Esseldeurs2026NatAs..10..124E} and
$R_\star\!\sim\!2.5$~au for V~Hya
\citep{Planquart2024A&A...682A.143P}, Eq.~8 of
\citet{Beguin_2024} predicts characteristic photocentre
displacements of $\sigma_{\rm conv}\!\sim\!0.38$~mas for
$\pi^1$~Gru and $\sigma_{\rm conv}\!\sim\!0.96$~mas for V~Hya.

\begin{table*}[htp]
	\centering
	\caption{Adopted input parameters and retrieved parameters for the  SS and VIMO models fitted to the simulated $\pi^1$~Gru and V~Hya Gaia DR5 epoch astrometry including red noise in the retrieval.  
}
	\label{tab:input_retrieved_params}
	\resizebox{\textwidth}{!}{%
		\begin{tabular}{l|c|ccc|c|cc}
			\hline\hline
			Parameter 
			& $\pi^1$~Gru input
			& \multicolumn{3}{c|}{$\pi^1$~Gru retrieved}
			& V~Hya input
			& \multicolumn{2}{c}{V~Hya retrieved} \\
			\cline{3-5}\cline{7-8}
			& 
			& SS & VIMO ($F_2$ known) & VIMO ($F_2$ unknown)
			&
			& SS & VIMO ($F_2$ known) \\
			\hline
			
			$\Delta \alpha^\star$
			& 0.000
			& $0.769 \pm 0.785$
			& $0.136 \pm 0.239$
			& $0.127 \pm 0.235$
			& 0.000
			& $-2.048 \pm 0.766$
			& $0.03 \pm 1.81$ \\
			
			$\Delta \delta$
			& 0.000
			& $-0.026 \pm 0.917$
			& $0.099 \pm 0.286$
			& $0.133 \pm 0.287$
			& 0.000
			& $4.364 \pm 0.752$
			& $1.93 \pm 2.46$ \\
			
			$\mu_{\alpha^\star}$ [mas\,a$^{-1}$]
			& 31.39
			& $34.438 \pm 0.261$
			& $31.083 \pm 0.233$
			& $31.108 \pm 0.228$
			& $-14.497$
			& $-12.471 \pm 0.225$
			& $-14.78 \pm 1.66$ \\
			
			$\mu_\delta$ [mas\,a$^{-1}$]
			& $-18.74$
			& $-20.349 \pm 0.238$
			& $-18.499 \pm 0.143$
			& $-18.523 \pm 0.140$
			& $-2.695$
			& $-1.283 \pm 0.213$
			& $-2.091 \pm 0.808$ \\
			
			$\varpi$ [mas]
			& 5.804
			& $6.081 \pm 0.226$
			& $5.842 \pm 0.070$
			& $5.840 \pm 0.070$
			& 2.3095
			& $2.566 \pm 0.235$
			& $2.453 \pm 0.173$ \\
			
			\hline
			
			$P$ [d]
			& 4337.76
			& --
			& $4505 \pm 119$
			& $4492 \pm 115$
			& 6373.61
			& --
			& $6371 \pm 3278$ \\
			
			$T_0$ [yr]
			& 2016.723
			& --
			& $2032.40 \pm 0.49^{(b)}$
			& $2032.32 \pm 0.51^{(b)}$
			& 2019.546
			& --
			& $2038.9 \pm 4.1^{(b)}$ \\
			
			$a$ [mas]
			& 37.48
			& --
			& --
			& --
			& 25.87
			& --
			& -- \\
			
			$\overline{\alpha}$ [mas]
			& 19.138
			& --
			& $20.176 \pm 0.808$
			& $20.069 \pm 0.787$
			& 13.415
			& --
			& $13.29 \pm 6.63$ \\
			
			$e$
			& 0.0
			& --
			& $0.033 \pm 0.023$
			& $0.032 \pm 0.022$
			& 0.024
			& --
			& $0.111 \pm 0.232$ \\
			
			$\omega$ [$^\circ$]
			& 0.0
			& --
			& $273.8 \pm 13.3^{(b)}$
			& $273.2 \pm 13.9^{(b)}$
			& 343
			& --
			& $233.4 \pm 98.7^{(b)}$ \\
			
			$i$ [$^\circ$]
			& 14.04
			& --
			& $19.67 \pm 3.62$
			& $19.40 \pm 3.61$
			& 37.7
			& --
			& $40.7 \pm 15.0$ \\
			
			$\Omega$ [$^\circ$]
			& 111.41
			& --
			& $118.96 \pm 5.63$
			& $118.37 \pm 5.73$
			& 159.7
			& --
			& $135.3 \pm 18.4$ \\
			
			$A_f$
			& 0.0037
			& --
			& $0.0421 \pm 0.0215$
			& $1.223 \pm 0.569$
			& 0.1112
			& --
			& $0.1083 \pm 0.0133$ \\
			
			\hline
			
			$M_1$ [$M_\odot$]
			& 0.929 & -- & -- & --
			& 1.93 & -- & -- \\
			
			$q$
			& 1.051 & -- & -- & --
			& 1.36 & -- & -- \\
			
			$P_{\rm puls}$ [d]
			& 195.5 & -- & -- & --
			& 530 & -- & -- \\
			
			$T_{0,\rm puls}$ [yr]
			& 1982.08 & -- & -- & --
			& 1995.70  & -- & -- \\
			
			$\Gamma$
			& 0.0 & -- & -- & --
			& $-0.68$ & -- & -- \\
			
			$G_{\rm max}$ [mag]
			& 3.253 & -- & -- & --
			& 5.08 & -- & -- \\
			
			$G_{\rm min}$ [mag]
			& 3.727 & -- & -- & --
			& 7.0 & -- & -- \\
			
			$G_{\rm main}$ [mag]
			& 3.5926 & -- & -- & --
			& 5.9249 & -- & -- \\
			
			$G_{\rm comp}$ [mag]
			& 10.3 & -- & -- & --
			& 8.8 & -- & -- \\
			
			$R_\star$ [au]
			& 1.65 & -- & -- & --
			& 2.5 & -- & -- \\
			
			$\tau_{\rm conv}$ [d]
			& 30 & -- & -- & --
			& 30 & -- & -- \\
			
			$\overline{\beta}$
			& 0.0018 & -- & -- & --
			& 0.05766 & -- & -- \\
			
			\hline
			
			$\sigma_{\rm jit}$ [mas]
			& 0.05
			& 0.00001
			& 0.00001
			& 0.00001
			& 0.05
			& 0.00001
			& 0.00001 \\
			
			$\sigma_{\rm red,\alpha^\star}$ [mas]
			& 0.38
			& $2.00^{(a)}$
			& $0.34$
			& $0.34$
			& 0.96
			& $2.00^{(a)}$
			& $0.85$ \\
			
			$\tau_{\rm red,\alpha^\star}$ [d]
			& 30
			& $300^{(a)}$
			& $20$
			& $20$
			& 30
			& $300^{(a)}$
			& $57$ \\
			
			$\sigma_{\rm red,\delta}$ [mas]
			& 0.38
			& $2.00^{(a)}$
			& $0.31$
			& $0.30$
			& 0.96
			& $2.00^{(a)}$
			& $0.99$ \\
			
			$\tau_{\rm red,\delta}$ [d]
			& 30
			& $300^{(a)}$
			& $20$
			& $20$
			& 30
			& $289$
			& $49$ \\
			\hline
			
			$\chi^2_{\rm red,gen}$ 
			& -- & 0.751 & 0.0126 & 0.0126
			& -- & 0.18 & 0.01 \\
			
			obj 
			& -- & $-1174$ & $-2444$ & $-2442$
			& -- & $-5870$ & $-6132$ \\
			
			DoF & -- & 1437 & 1429 & 1429 & -- & 1881 & 1873 \\

			\hline
		\end{tabular}%
	}
	\tablefoot{Input orbital and stellar parameters for
		$\pi^1$~Gru are primarily based on \citet{Esseldeurs2026NatAs..10..124E},
		while the adopted V~Hya parameters are based on
		\citet{Planquart2024A&A...682A.143P}. 
		For clarity, estimator symbols (e.g.\ hats) are omitted throughout
		the table; all quoted quantities in the columns listing the retrieved parameters correspond to fitted parameter
		estimates. The quoted uncertainties are formal $1\sigma$ errors including a correlated red-noise covariance model.
		For $\pi^1$~Gru, two VIMO configurations are shown: one where the companion
		flux $F_2$ is assumed to be known in the construction of the variability
		term entering the VIMO fit, and one where the companion flux is assumed
		to be unknown.
		For V~Hya, only the single-star and VIMO retrievals with fixed companion
		flux were considered.
		In the $\pi^1$~Gru unknown-flux VIMO case, the retrieved amplitude parameter $A_f$
		should be interpreted in the context of the reconstructed variability term
		$\tilde{f}(t)$ rather than the true variability term $f(t)$. The last row lists the number of degrees of freedom (DoF) for each fitting procedure. \\
		$^{(a)}$ Convergence toward imposed parameter boundary.
		$^{(b)}$ Because the simulated $\pi^1$~Gru and V~Hya orbits are nearly circular, the argument of periastron $\omega$ and time of periastron passage $T_0$ are intrinsically ill constrained.
	}
\end{table*}

\begin{figure*}[htp]
	\sidecaption
	\includegraphics[width=12cm]{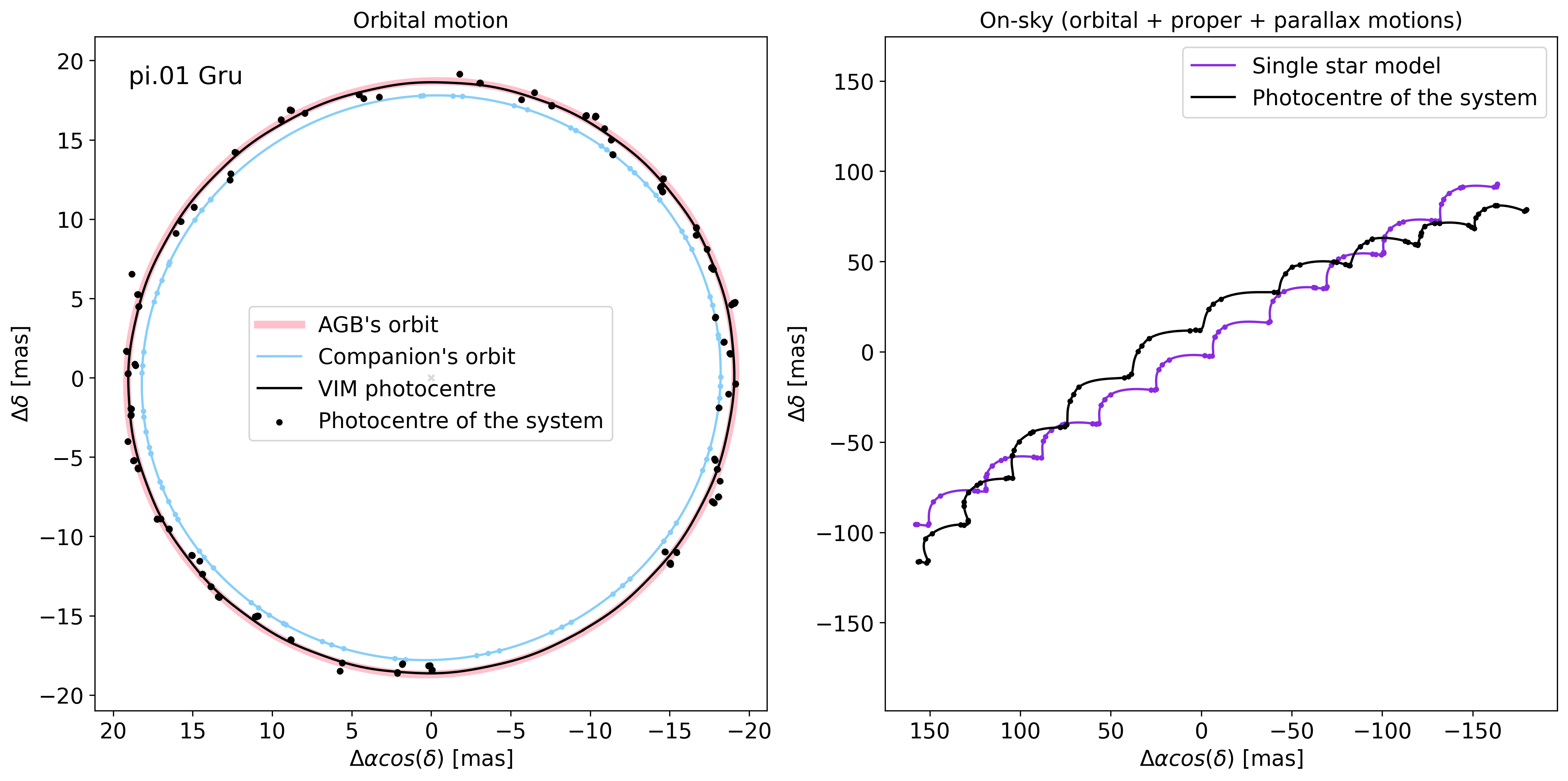}
	\caption{Forward simulations of the astrometric motion of the $\pi^1$~Gru A-C system.
		The left panel shows the orbital motion around the center of mass of $\pi^1$~Gru A (pink) and its companion $\pi^1$~Gru C (blue). The corresponding photocentric orbit including only the pulsation-induced VIM effect is shown as a full black line. The black dots correspond to the photocentric orbit over the DR5 time span including both the VIM effect and convection-induced photocentric shifts. Right panel: Single-star sky path (in purple) with the same proper motion and parallactic motion as $\pi^1$~Gru compared with the full system description including binary, variability, and convection effects (in black).}
	\label{fig:forward_simulation_pi1gru}
\end{figure*}

\begin{figure*}[htp]
	\sidecaption
	\includegraphics[width=12cm]{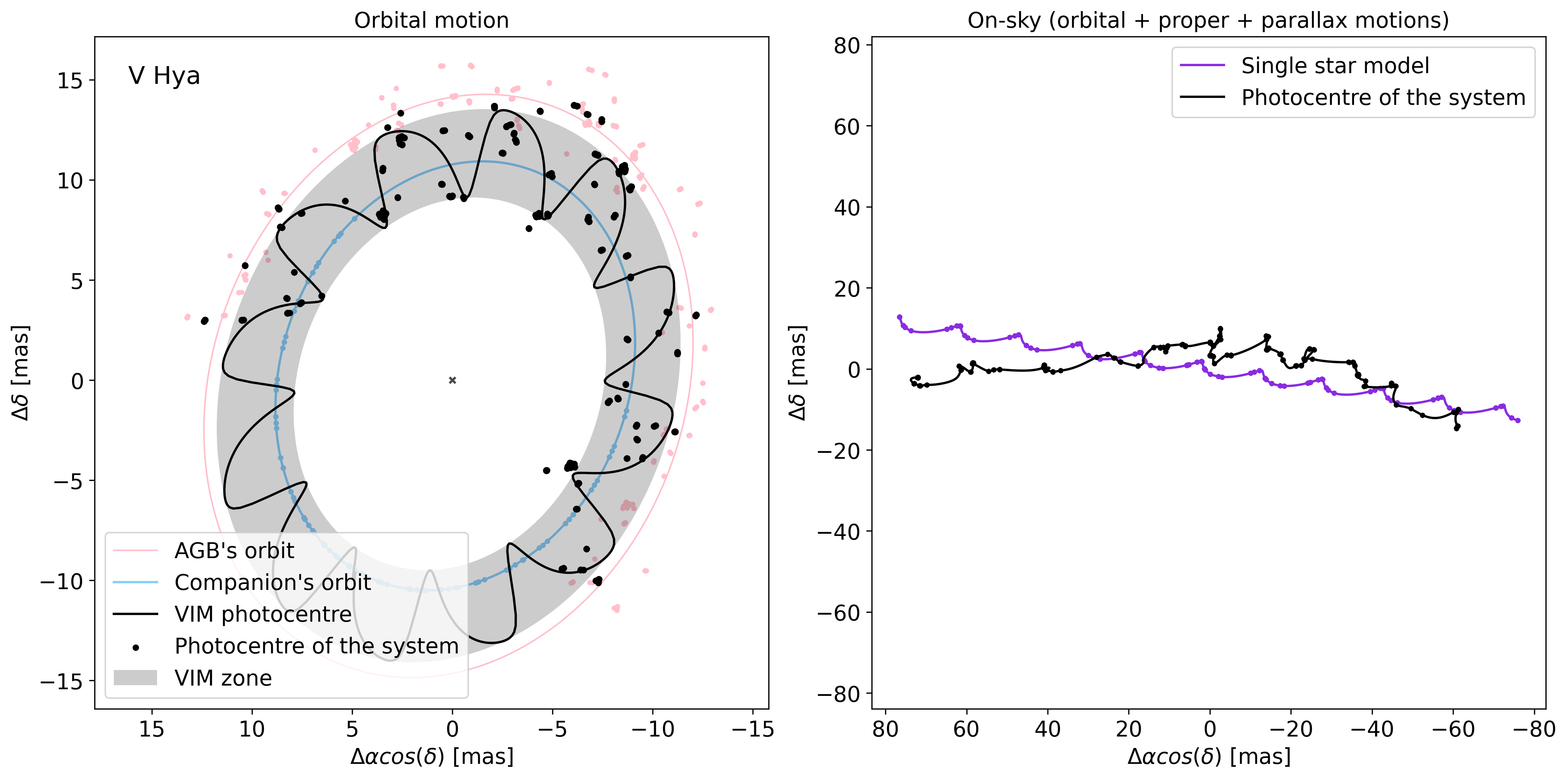}
	\caption{See Fig.~\ref{fig:forward_simulation_pi1gru}, but here for V~Hya. The pink dots show the effect of convection on the photocentre of the primary AGB star. The grey zone indicates the pulsation-induced VIM amplitude.}
	\label{fig:forward_simulation_VHya}
\end{figure*}

A summary of all input parameters is given in Table~\ref{tab:input_retrieved_params}, with the (on-sky) forward simulations shown in Fig.~\ref{fig:forward_simulation_pi1gru} and Fig.~\ref{fig:forward_simulation_VHya} and corresponding simulated Gaia epoch astrometry in Figs.~\ref{fig:pi1gru_retrieved_known_Vcomp}\,--\,\ref{fig:VHya_retrieved_known_Vcomp}.
After validating the VIMO methodology in the DR5 configuration (Sect.~\ref{Sec:validation_AGB}\,--\,\ref{Sec:validation_AGB_VHya}),
we investigate how the retrieved astrometric parameters, in
particular the parallax and proper motion, may differ between
the various Gaia data releases (DR2, DR3, DR4, and DR5; see Sect.~\ref{Sec:Discussion}) \footnote{\label{footnote_timesampling}Our simulations additionally show that some retrieved parameters can depend sensitively on the exact temporal sampling of the observations. Since the actual Gaia epoch timestamps are not yet publicly available, we relied on the observation epochs predicted by the Gaia Observation Schedule Tool (GOST). These predicted timestamps are expected to differ slightly from the true Gaia observing times, implying that some realization-dependent details of the simulated solutions may differ from the final Gaia epoch astrometry.}.

We first validate the retrieval framework on $\pi^1$~Gru, for which
the VIM effect remains
small ($d_{\rm var}\sim0.03$~mas). In this system, the expected
convection-induced photocentre variability is larger than the
pulsation-induced VIM signal, such that the astrometric variability
is dominated by the combination of Keplerian orbital motion and
stochastic convection-induced photocentre jitter. $\pi^1$~Gru
therefore provides a comparatively clean benchmark for validating
the red-noise retrieval framework in the presence of an unresolved
Keplerian orbit.

As an additional validation of the red-noise retrieval framework,
we also performed a separate $\pi^1$~Gru simulation in which the
convection-induced photocentre shifts were taken directly from
the three-dimensional radiation-hydrodynamic model
\texttt{st29gm04n001} of \citet{Beguin_2024}, rather than from
an OU realization. This test allows us to assess whether the
VIMO retrieval remains robust when the injected convective
signal follows a physically motivated hydrodynamical time series
instead of the same stochastic model used in the covariance
description.

We subsequently apply the methodology to the far more challenging
V~Hya system, where the pulsation-induced photocentre variability
($d_{\rm var}\!\sim\!5.7$~mas)  reaches nearly 40\% of the primary’s orbital astrometric displacement about the barycentre.
In that regime, the retrieval
must simultaneously disentangle Keplerian orbital motion,
pulsation-induced VIM variability, and stochastic
convection-induced photocentre displacements.

\subsection{Recovery of the simulated \texorpdfstring{$\pi^1$}{pi1}~Gru parameters}\label{Sec:validation_AGB}

To assess the performance of the different astrometric retrieval
approaches, we fitted the simulated Gaia DR5 epoch astrometry of
$\pi^1$~Gru (see Fig.~\ref{fig:forward_simulation_pi1gru}) with several increasingly sophisticated models.
First, we considered the classical SS, VIMF, VIML,
and VIMA models, without including correlated
red noise in the retrieval. These fits provide a direct comparison with the traditional variability-induced mover formalisms more commonly applied to Gaia and Hipparcos astrometry. The retrieved parameters are  summarized in Table~\ref{tab:pi1gru_vim_no_rednoise}.

\begin{figure*}[htp]
	\centering
	\includegraphics[width=\textwidth]{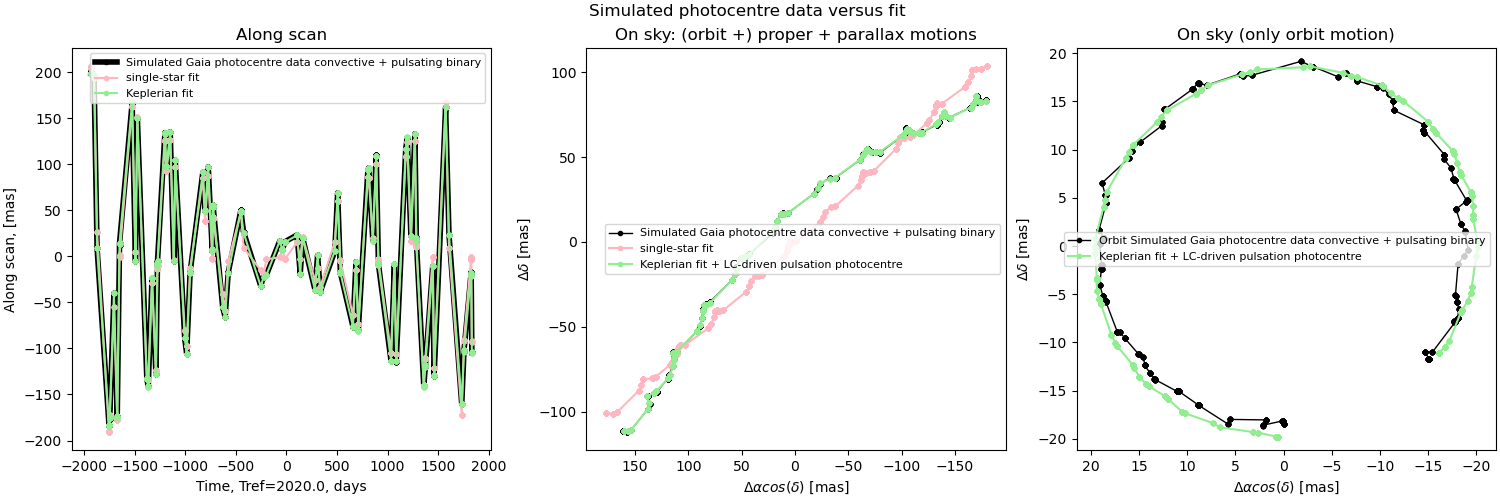}
	\caption{
		Comparison between the simulated Gaia DR5 epoch astrometry of the
		$\pi^1$~Gru system and the retrieved astrometric models in the case
		where the companion flux is assumed to be known in the construction of
		the variability term entering the VIMO fit; see Table~\ref{tab:input_retrieved_params}.
		The left panel shows the along-scan Gaia epoch astrometry as a
		function of time.
		The black symbols correspond to the simulated photocentre data,
		including binary orbital motion, pulsation-induced variability, and
		convection-induced photocentre shifts.
		The pink curve shows the best-fit single-star solution, while the
		green curve shows the best-fit VIMO model including the
		light-curve-driven photocentre variability.
		The middle panel compares the reconstructed sky trajectories,
		including (orbital motion,)  proper motion, and parallax. The black data points are largely hidden behind the green curve owing to the good agreement between the two.
		The right panel isolates the orbital component alone.}
	\label{fig:pi1gru_retrieved_known_Vcomp}
\end{figure*}

\begin{figure}[htp]
	\centering
	\includegraphics[width=\columnwidth]{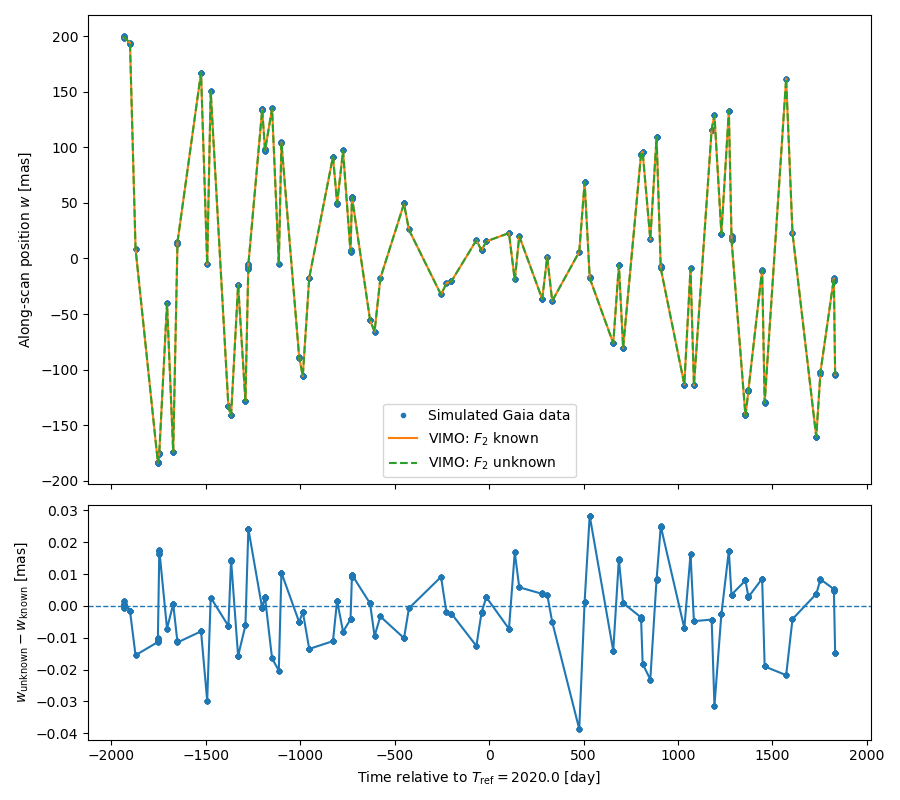}
	\caption{Comparison between the best-fitting VIMO models for $\pi^1$~Gru obtained when the companion flux is assumed to be known and when it is treated as unknown. Top: Simulated Gaia DR5 along-scan epoch astrometry together with the two best-fitting VIMO solutions. The two model predictions are nearly indistinguishable on the scale of the full astrometric signal. Bottom: Difference between the predicted along-scan positions, $w_{\rm unknown} - w_{\rm known}$, at the simulated observing epochs. The rms difference between the two solutions is 0.012\,mas, with a maximum absolute difference of 0.039\,mas. Despite the substantially different fitted values of $A_f$, the recovered orbital and astrometric solutions therefore remain essentially unchanged.}
	\label{fig:pi1gru_comparison}
\end{figure}

\begin{figure*}[htp]
	\centering
	\includegraphics[width=\textwidth]{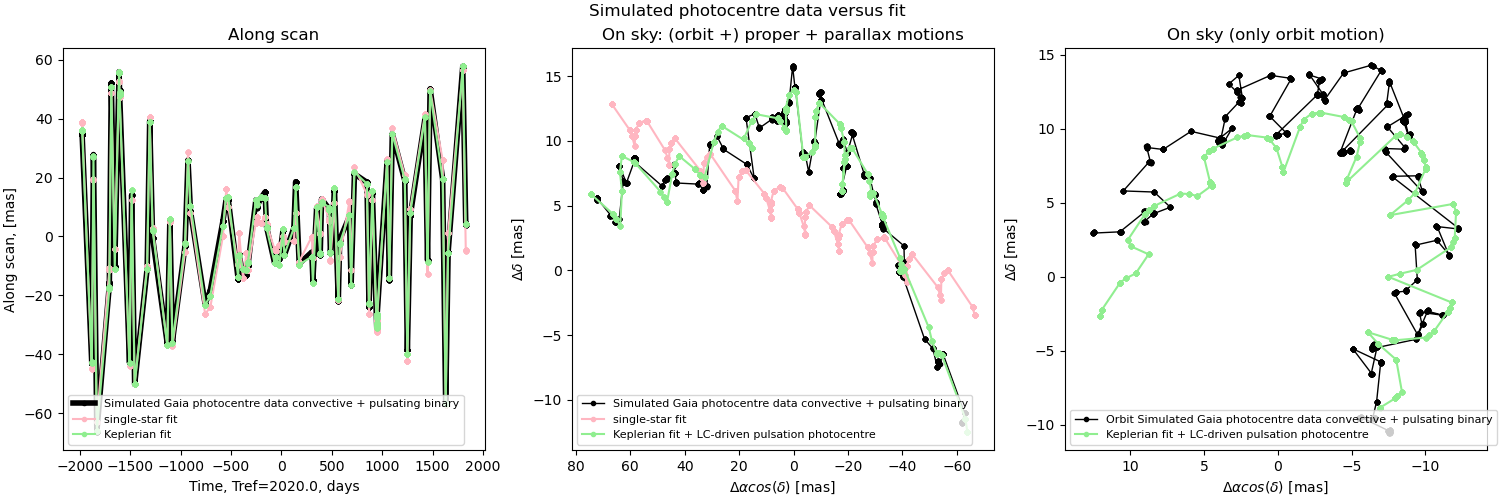}
	\caption{Same as Fig.~\ref{fig:pi1gru_retrieved_known_Vcomp}, but here for V~Hya. Retrieved model parameters are listed in  Table~\ref{tab:input_retrieved_params}.}
	\label{fig:VHya_retrieved_known_Vcomp}
\end{figure*}

Subsequently, we applied the combined VIMO+CIM retrieval framework,
which combines Keplerian orbital motion with pulsation-induced
VIM and includes a correlated-noise covariance component to
account for convection-induced photocentre motion; henceforth called the red-noise VIMO retrieval.
Two retrieval scenarios were considered: (i)~a fit assuming that the companion flux is known, and (ii)~a fit in which the companion flux is treated as unknown, such that the variability term is reconstructed through the modified flux function $\tilde{f}(t)$. The retrieved parameters of the red-noise SS and VIMO retrievals can be found in Table~\ref{tab:input_retrieved_params}. Figure~\ref{fig:pi1gru_retrieved_known_Vcomp} shows the resulting fit for the known-flux case, while Fig.~\ref{fig:pi1gru_comparison} directly compares the two VIMO solutions. The fitted along-scan trajectories are nearly indistinguishable: their rms difference is only 0.012 mas, with a maximum absolute difference of 0.039 mas. This explains why the two solutions would appear essentially identical when plotted on the full astrometric scale.

The comparison between the input parameters and the retrieved parameters
from the different astrometric models 
illustrates both the limitations of simplified VIM descriptions
and the importance of explicitly accounting for Keplerian orbital
motion and correlated photocentre variability in AGB systems.
The simplest SS solution (see Table~\ref{tab:pi1gru_vim_no_rednoise}) produces strongly biased astrometric parameters. In particular,
the retrieved proper motion components differ substantially from
the barycentric input values.
The extremely large reduced $\chi^2$ value (of 467) further confirms that a single-star model is unable to reproduce the simulated Gaia
epoch astrometry of $\pi^1$~Gru.
	
The VIMF and VIML models partially improve the fit by
introducing photocentre displacements correlated with the
photometric variability. However, these models still rely on
low-order approximations of the orbital motion. Significant biases remain in the recovered proper motion and parallax values. The VIML model performs slightly better than the VIMF solution because it allows for a linear evolution of the photocentre displacement, partially capturing the long-term orbital drift over the Gaia observing window.

The VIMA model provides a substantially improved description
of the simulated data. By including acceleration terms, the model
captures part of the orbital curvature and recovers astrometric
parameters that are significantly closer to the input values.
The recovered parallax agrees well with the simulated value. Nevertheless, the formal fit quality remains unsatisfactory because the VIMA formalism still represents only a low-order approximation of the true Keplerian motion. 

An additional indication that the white-noise VIM prescriptions do
not provide an adequate description of the $\pi^1$~Gru astrometry
is the unrealistically large value of the inferred photocentre
offset $D = \sqrt{D_\alpha^2 + D_\delta^2}$, which nominally represents the angular separation between the
variable primary star and the system photocentre (see Paper~I). In the absence
of strong pulsation-induced variability, such large inferred
photocentre offsets are physically implausible for the
$\pi^1$~Gru system and indicate that the deterministic VIM model
is attempting to absorb unresolved orbital motion and stochastic convection-induced astrometric
variability that is not represented adequately within the
white-noise framework. This interpretation is further supported
by the correspondingly large values of $\chi^2_{\rm red}$ obtained
for these fits.

The red-noise single-star solution fails to reproduce the on-sky photocentre trajectory (see the pink curve in the middle panel of Fig.~\ref{fig:pi1gru_retrieved_known_Vcomp}), whereas the red-noise VIMO solution successfully recovers the dominant orbital motion despite the presence of both pulsation-induced variability and stochastic convection-induced photocentre shifts (green curve in the same panel). The remaining small differences between the simulated and recovered photocentre trajectories can be attributed to the stochastic nature of the convection-induced photocentre variability together with the finite temporal sampling of the Gaia observations. Overall, the red-noise VIMO retrieval framework recovers the simulated system parameters accurately (see Table~\ref{tab:input_retrieved_params}). In particular, the retrieved orbital period, inclination, and longitude of the ascending node are all consistent with the input values within their formal uncertainties. The recovered photocentric semi-major axis, $\overline{\alpha}$, is likewise in good agreement with the injected photocentric orbit size. Furthermore, the fitted red-noise amplitudes and correlation timescales remain physically plausible and are consistent with the stochastic convection-induced photocentre variability injected into the forward simulations.

We additionally explored the sensitivity of the retrievals to
the adopted OU-process correlation timescale by performing
forward simulations for a range of $\tau_{\rm conv}$ values.
Although the detailed realization of the stochastic
photocentre variability changes with the adopted correlation
timescale, the recovered orbital and astrometric parameters
remain stable within their formal uncertainties.

The comparison between the retrievals assuming a known or an unknown companion flux further illustrates two distinct effects. First, in the absence of convection-induced photocentre variability, tests performed with the same VIMO retrieval framework show that, when the companion flux is known, the fitted amplitude $\hat{A}_f$ correctly recovers the injected physical variability amplitude $A_f$, as expected from Eq.~(14) for $\tilde{f}(t)=f(t)$. Second, when convection-induced photocentre variability is included, as in the simulations presented here, the fitted VIMO amplitude can absorb part of the correlated CIM signal. This explains why the known-flux $\pi^1$~Gru retrieval does not recover the injected value of $A_f$ exactly, even though the companion flux is fixed. The effect is particularly noticeable for $\pi^1$~Gru because the pulsation-induced VIM signal is intrinsically small compared to the convection-induced photocentre excursions. When the companion flux is treated as unknown, an additional degeneracy is introduced: the fitted variability term must then be interpreted in the context of the modified flux function $\tilde{f}(t)$ rather than the true variability term $f(t)$ (see Sect.~\ref{Sec:VIMO_retrieval}). Although the retrieved value of $A_f$ can therefore differ substantially from the physical input value, the orbital parameters themselves remain well recovered.

We performed an additional robustness test for $\pi^1$~Gru in
which the convection-induced photocentre motion was injected
from the hydrodynamical simulation \texttt{st29gm04n001}
instead of from an OU process (see Fig.~\ref{fig:pi1gru_retrieved_unknown_Vcomp_conv_Chiav}). 
The red-noise single-star fit
again absorbs part of the correlated photocentre variability but
does not recover the barycentric proper motion: it yields
$\varpi = 5.819 \pm 0.225$~mas,
$\mu_{\alpha^\star}=34.340 \pm 0.254$~mas\,a$^{-1}$, and
$\mu_\delta=-20.310 \pm 0.228$~mas\,a$^{-1}$.
In contrast, the corresponding full VIMO fit 
recovers the orbital solution well, with
$P=4558\pm349$~d,
$\overline{\alpha}=21.38\pm2.33$~mas,
$i=22.4\pm7.3^\circ$, and
$\Omega=143.7\pm20.4^\circ$.
The recovered parallax,
$\varpi=5.677\pm0.133$~mas, remains consistent with the input
value within the formal uncertainties.
This test is important because the injected convection signal no
longer follows the same OU prescription used in the retrieval
covariance model. The successful recovery of the dominant
Keplerian parameters therefore indicates that the full VIMO
framework is not merely tuned to OU-generated noise, but remains
robust for a more realistic hydrodynamical description of
surface-convection photocentre variability.

\subsection{Recovery of the simulated V~Hya parameters}\label{Sec:validation_AGB_VHya}

We next applied the methodology to V~Hya, for which the VIM effect is substantially stronger than in $\pi^1$~Gru. The retrieved parameters for the SS, VIMF, VIML, and VIMA fits including only white noise are listed in Table~\ref{tab:vhya_vim_no_rednoise}, while the parameters of the red-noise VIMO fit are given in Table~\ref{tab:input_retrieved_params}. Overall, the same trends are observed as for $\pi^1$~Gru: models of increasing complexity progressively improve the recovery of the astrometric parameters. 

The white-noise SS, VIMF, and VIML fits yield substantially
larger reduced $\chi^2_{\rm red}$ values for V~Hya than for
$\pi^1$~Gru, reflecting the much stronger photocentre
variability and the larger VIM signal in the system
($d_{\rm var}\sim5.7$~mas for V~Hya compared to only
$\sim0.03$~mas for $\pi^1$~Gru). The retrieved parallaxes of
the white-noise SS and simplified VIM models are strongly
biased and fail to recover the input parallax within the formal
uncertainties (Table~\ref{tab:vhya_vim_no_rednoise}).

In contrast, the VIMA model produces a significantly lower
$\chi^2_{\rm red}$ for V~Hya. This apparent improvement arises
primarily because the long orbital period of V~Hya
($P\!\sim\!17.5$~yr) implies that Gaia samples only
$\sim$\,60\% of the orbit, causing the unresolved orbital motion
to resemble a low-order astrometric drift that can be partially
absorbed by the acceleration terms of the VIMA prescription.

Most importantly, the  red-noise VIMO framework
successfully recovers all major orbital parameters, including
the photocentre semi-major axis $\overline{\alpha}$ and the
variability amplitude parameter $A_f$, despite the strong
pulsation-induced photocentre variability and the incomplete
orbital phase coverage. The recovered red-noise hyperparameters
remain physically consistent with the injected
convection-induced photocentre variability.

\subsection{Robustness of the VIMO retrieval framework}

Overall, these simulations demonstrate that the retrieval framework remains robust even in the challenging regime where both pulsation-induced VIM and convection-induced photocentre variability contribute significantly to the observed astrometric signal. Although the convection-induced photocentre motion introduces temporally correlated residuals and the pulsation-induced photocentre excursions reach amplitudes comparable to a substantial fraction of the orbital signal, the combined VIMO and red-noise retrieval is still able to recover the injected orbital parameters reliably. This remains true not only for stochastic Ornstein–Uhlenbeck realizations of the convection signal, but also when the injected photocentre variability is taken directly from the three-dimensional radiation-hydrodynamic simulations. These results indicate that Gaia DR4/DR5 epoch astrometry can provide reliable orbital solutions for unresolved nearby AGB binaries even in the presence of strong intrinsic photocentre variability.

\section{Discussion} \label{Sec:Discussion}

\subsection{Biases in Gaia DR2, DR3, and DR4 astrometric solutions} \label{Sec:biases}

While the previous subsection demonstrated that the full orbital and astrometric parameters of the simulated AGB binary systems can in principle be recovered from Gaia-like epoch astrometry using the red-noise VIMO framework, the currently available Gaia catalogue solutions are based on substantially shorter observing baselines and simpler astrometric models. As a consequence, unresolved orbital motion, pulsation-induced photocentre shifts, and convection-induced variability may bias the catalogued parallaxes and proper motions.

To investigate these effects, we repeated the forward
simulations for observing baselines representative of Gaia
DR2, DR3, and DR4, and fitted the resulting epoch astrometry
with a single-star astrometric model. This allows us to
quantify how the inferred catalogue parameters depend on the
mission duration and orbital phase coverage.
The retrieved single-star solutions are summarized in
Tables~\ref{tab:pi1gru_dr2dr3dr4}
and~\ref{tab:vhya_dr2dr3dr4}. In all cases, the simulated
epoch astrometry was fitted without accounting for unresolved
orbital motion, pulsation-induced VIM, or convection-induced
photocentre variability.

The resulting astrometric parameters vary substantially between
the different Gaia releases, illustrating how strongly the inferred
catalogue solution depends on the temporal sampling and orbital
phase coverage. In particular, the retrieved proper motion values
show large systematic differences between DR2, DR3, and DR4.
The DR2 and DR3 solutions recover strongly biased declination
proper motions, while the DR4 solution begins to approach the
true (barycentric) motion as a larger fraction of the orbit becomes
sampled. Nevertheless, even DR4 still produces substantially
biased astrometric parameters when the unresolved orbital motion
is ignored. The retrieved parallaxes also vary between the different releases at the level of several tenths of a milliarcsecond ($\pi^1$~Gru) up to $\sim$1~mas for V~Hya. 

For comparison, the $\pi^1$~Gru Gaia DR3 catalogue reports
$\varpi = 6.188 \pm 0.199$~mas,
$\mu_{\alpha\star} = 31.106 \pm 0.063$~mas\,a$^{-1}$,
and $\mu_{\delta} = -10.338 \pm 0.121$~mas\,a$^{-1}$.
These values differ moderately from our simulated DR3 retrievals.
Part of this discrepancy is expected to arise from differences in
the exact Gaia temporal sampling adopted in the simulations
(see footnote~\ref{footnote_timesampling}).
In addition, the forward simulations necessarily rely on
simplified prescriptions for both the pulsation and convection
signals. In particular, the convection-induced photocentre
variability is stochastic and therefore cannot be predicted
deterministically for the real system, while the pulsation
variability of $\pi^1$~Gru is approximated here by an idealized
periodic light-curve model that does not reproduce the full
irregularity observed in the Gaia epoch photometry.

The retrievals further demonstrate that the (DR5) recovered parallaxes of $\pi^1$~Gru and V~Hya 
depend  on the adopted astrometric model prescription and on whether correlated red noise is included in the retrieval; see 
Table~\ref{tab:input_retrieved_params} and
Tables~\ref{tab:pi1gru_vim_no_rednoise}\,--\,\ref{tab:vhya_vim_no_rednoise}.
For $\pi^1$~Gru, the input parallax adopted in the forward simulations, $\varpi = 5.804$~mas, falls outside the formal $1\sigma$ uncertainties of the white-noise VIMF, VIML and VIMA fits, whereas the SS model recovers values consistent with the input parallax. For V~Hya, none of the white-noise SS or VIM models recover the input parallax of $\varpi = 2.3095$~mas within their formal uncertainties. 
The comparison between the simulated DR2, DR3, DR4, and DR5 white-noise single-star solutions (see
Tables~\ref{tab:pi1gru_dr2dr3dr4}
and~\ref{tab:vhya_dr2dr3dr4}) further shows that the
retrieved parallaxes are never consistent with the input values
within their formal uncertainties.
In contrast, for both systems, the parallax is accurately recovered when correlated red noise is included in the SS and VIMO fits. 

\subsection{Impact of correlated red noise on Gaia DR4 astrometric retrievals}

To investigate whether the inclusion of a red-noise covariance matrix can already improve the interpretation of Gaia DR4 epoch astrometry, we repeated the DR4 retrievals using both the red-noise SS and  VIMO frameworks. The retrieved parameters are summarized in Table~\ref{tab:pi1gru_vhya_dr4_rednoise} and can be compared directly to the white-noise DR4 retrievals listed in Tables~\ref{tab:pi1gru_dr2dr3dr4} and \ref{tab:vhya_dr2dr3dr4}, as well as to the input parameters adopted in the forward simulations (Table~\ref{tab:input_retrieved_params}).

The DR4 retrievals confirm that the Gaia observing baseline remains too short to  constrain the orbital motion of either $\pi^1$~Gru or V~Hya. 
Even for $\pi^1$~Gru, where the orbital constraints are significantly tighter, the retrieved orbital period of $\sim$\,2080~d still differs substantially from the input period of 4338~d.

Furthermore, the recovered proper motions remain significantly biased with respect to the  input values. For $\pi^1$~Gru, the input proper motions are $\mu_{\alpha^\star}=31.39$ mas\,a$^{-1}$ and $\mu_\delta=-18.74$ mas\,a$^{-1}$, whereas the DR4 red-noise VIMO retrieval yields $\mu_{\alpha^\star}=35.83 \pm 0.18$ mas\,a$^{-1}$ and $\mu_\delta=-14.36 \pm 0.72$ mas\,a$^{-1}$. Similarly, for V~Hya, the recovered DR4 VIMO proper motions remain offset from the simulated barycentric values by several tenths to several mas\,a$^{-1}$. These biases arise because Gaia DR4 still samples only a limited fraction of the orbital motion, such that long-period orbital curvature is partially absorbed into the astrometric proper-motion terms.

Nevertheless, an important result emerges from the comparison between the white-noise and red-noise retrievals: including correlated red noise substantially improves the recovery of the parallax, even when using only a single-star astrometric model. This effect is particularly evident when comparing the DR4 white-noise single-star solutions (Tables~\ref{tab:pi1gru_dr2dr3dr4} and \ref{tab:vhya_dr2dr3dr4}) with the corresponding red-noise SS retrievals in Table~\ref{tab:pi1gru_vhya_dr4_rednoise}.

For $\pi^1$~Gru, the input parallax adopted in the forward simulations is $\varpi = 5.804$ mas. The DR4 white-noise single-star retrieval yielded $\varpi = 5.041 \pm 0.284$ mas, corresponding to a bias of approximately $-0.76$ mas. After including correlated red noise in the DR4 SS retrieval, the recovered parallax becomes $\varpi = 5.758 \pm 0.316$ mas, in excellent agreement with the input value. A similar improvement is observed for V~Hya. The input parallax is $\varpi = 2.3095$~mas, while the
white-noise DR4 SS retrieval yields
$\varpi = 2.126 \pm 0.118$~mas.
Including correlated red noise shifts the retrieved value to
$\varpi = 2.534 \pm 0.279$~mas, which remains statistically
consistent with the input parallax within the formal
uncertainties.

The improvement in parallax recovery demonstrates that part of the astrometric bias introduced by pulsations and convection-induced photocentre variability can already be mitigated through an appropriate covariance treatment, even when the orbital motion itself remains only partially constrained. Physically, the red-noise covariance model prevents the low-frequency stochastic photocentre excursions from being absorbed into the parallax solution. In contrast, when only white noise is assumed, the retrieval attempts to reproduce the correlated photocentre variability through distortions of the astrometric parameters themselves, thereby biasing both the parallax and proper motion estimates.

These results suggest that incorporating correlated astrometric noise models into future Gaia non-single-star analyses may already provide significantly improved parallaxes for intrinsically variable evolved stars in DR4, even before the full orbital motion becomes recoverable in DR5-like observing baselines.

\section{Conclusions} \label{Sec:conclusions}

We investigated how unresolved orbital motion,
pulsation-induced variability, and convection-induced
photocentre displacements affect the Gaia astrometry of nearby
AGB binary systems. Building upon the general VIM framework
introduced in Paper~I, we developed a retrieval methodology
capable of simultaneously modelling Keplerian photocentre
motion, coherent pulsation-induced variability, and stochastic
convection-induced astrometric jitter through a correlated
red-noise covariance treatment.

The methodology was applied to the only two currently
well-characterised nearby AGB binary systems,
$\pi^1$~Gru and V~Hya, which probe two complementary
variability regimes. For $\pi^1$~Gru, the astrometric
variability is dominated primarily by the combination of
Keplerian orbital motion and convection-induce 
excursions strongly modulate the Keplerian photocentre orbit and become a dominant component of the observed astrometric variability.

Our simulations demonstrate that Gaia DR4 and especially Gaia
DR5 epoch astrometry should enable the recovery of accurate
orbital solutions and parallaxes for nearby unresolved AGB
binaries, provided that the photocentre variability is modeled
adequately. Simplified single-star or low-order white-noise
VIM models can yield substantially biased proper motions and
parallaxes, with the inferred astrometric parameters depending
sensitively on the Gaia observing baseline and orbital phase
coverage.

The retrieval experiments further show that convection-induced
photocentre variability does not fundamentally prevent the
astrometric characterisation of nearby AGB binaries.
Although stochastic surface structures introduce correlated
astrometric noise at the sub-milliarcsecond level,
incorporating a physically motivated covariance model allows
the dominant Keplerian signal to be recovered reliably. The
additional robustness test using convection-induced
photocentre variability extracted directly from a
three-dimensional radiation-hydrodynamic simulation further
demonstrates that the red-noise VIMO framework remains capable
of recovering the dominant orbital and astrometric parameters
even when the injected variability no longer follows the same
stochastic prescription adopted in the retrieval covariance
model.

An important implication of this work is that the correlated
red-noise treatment adopted here is not part of the current
Gaia NSS (Non-Single Star) processing framework. Present
and future Gaia NSS solutions therefore do not explicitly
account for stochastic convection-induced photocentre
variability, despite theoretical predictions and observational
evidence that such effects can reach sub-milliarcsecond
amplitudes in nearby evolved stars. Our simulations suggest
that incorporating physically motivated correlated-noise
treatments may substantially improve the astrometric
characterisation of intrinsically variable evolved binaries in
future extensions of the Gaia NSS framework.

At present, $\pi^1$~Gru and V~Hya remain the only two nearby
AGB binary systems with sufficiently complete independent
orbital constraints to permit a detailed validation of the red-noise
VIMO methodology. However, Gaia DR4 and especially Gaia
DR5 are expected to substantially increase the number of
well-characterised unresolved AGB binaries through the
combination of epoch astrometry, photometric variability,
and non-single-star orbital solutions. Extending the present
framework to larger samples will provide new opportunities
for binary characterization during late stellar
evolution and to improve the astrometric calibration of
long-period variable stars used as distance indicators for
older stellar populations.

\begin{acknowledgements}
	  LD acknowledges support from the
	  FWO grant G0B3823N, the  FWO grant G099720N, the KU Leuven C1 excellence grant MAESTRO
	  C16/17/007, and the KU Leuven Methusalem SOUL grant METH/24/012.
      This work received funding from ANR Unlock-pfactor (ANR-23-CE31-0009). In addition, this research is supported by the European Research Council (ERC) under the European Union’s Horizon 2020 research and innovation program (grant agreement No. 951549 - UniverScale). PK acknowledges support from the Polish-French Marie Skłodowska-Curie and Pierre Curie Science Prize awarded by the Foundation for Polish Science. AC acknowledges support from the French National Research Agency (ANR) funded project
      PEPPER (ANR-20-CE31-0002).  This work has made use of data from the European Space Agency (ESA) mission
      {\it Gaia} (\url{https://www.cosmos.esa.int/gaia}), processed by the {\it Gaia}
      Data Processing and Analysis Consortium (DPAC,
      \url{https://www.cosmos.esa.int/web/gaia/dpac/consortium}). Funding for the DPAC
      has been provided by national institutions, in particular the institutions
      participating in the {\it Gaia} Multilateral Agreement.
\end{acknowledgements}

\bibliographystyle{aa} 
\bibliography{bib}

@ARTICLE{kepmodel,
       author = {{Delisle}, J.-B. and {S{\'e}gransan}, D.},
        title = "{Analytical determination of orbital elements using Fourier analysis. II. Gaia astrometry and its combination with radial velocities}",
      journal = {\aap},
         year = 2022,
        month = nov,
       volume = {667},
          eid = {A172},
        pages = {A172},
          doi = {10.1051/0004-6361/202244018},
archivePrefix = {arXiv},
       eprint = {2209.13992},
 primaryClass = {astro-ph.EP},
       adsurl = {https://ui.adsabs.harvard.edu/abs/2022A&A...667A.172D}
}

@article{UhlenbeckOrnstein1930,
	author  = {Uhlenbeck, G. E. and Ornstein, L. S.},
	title   = {On the Theory of the Brownian Motion},
	journal = {Physical Review},
	volume  = {36},
	pages   = {823--841},
	year    = {1930},
	doi     = {10.1103/PhysRev.36.823}
}

@ARTICLE{Chiavassa2022A&A...661L...1C,
	author = {{Chiavassa}, A. and {Kudritzki}, R. and {Davies}, B. and {Freytag}, B. and {de Mink}, S.~E.},
	title = "{Probing red supergiant dynamics through photo-center displacements measured by Gaia}",
	journal = {\aap},
	year = 2022,
	month = may,
	volume = {661},
	eid = {L1},
	pages = {L1},
	doi = {10.1051/0004-6361/202243568},
	archivePrefix = {arXiv},
	eprint = {2205.05156},
	primaryClass = {astro-ph.SR},
	adsurl = {https://ui.adsabs.harvard.edu/abs/2022A&A...661L...1C}
}

@ARTICLE{Kochanek2023MNRAS.520.3510K,
	author = {{Kochanek}, C.~S.},
	title = "{A non-detection of red supergiant convection in Gaia}",
	journal = {\mnras},
	year = 2023,
	month = apr,
	volume = {520},
	number = {3},
	pages = {3510-3513},
	doi = {10.1093/mnras/stac2780},
	archivePrefix = {arXiv},
	eprint = {2206.09926},
	primaryClass = {astro-ph.SR},
	adsurl = {https://ui.adsabs.harvard.edu/abs/2023MNRAS.520.3510K}
}

@ARTICLE{PaperI,
author = {{Sivkova}, K. and {Kervella}, P. and {Gallenne}, A. and {Decin}, L. and {Nardetto}, N. and {M\'erand}, A. and {Apostolova}, B. and {Bailleul}, M.~C. and {Rathour}, R.~S. and {Kiviaho}, W. and {Bras}, G. and {Hocd\'e}, V. and {Breuval}, L. and {Vert}, C.},
title = "{Effect of binarity and variability on the astrometry of unresolved systems. I. Simulations and framework to interpret Gaia DR4 epoch astrometry}",
journal = {\aap},
year = 2026,
month = may,
volume = {subm.},
eid = {}
}

@ARTICLE{Huang2024ApJ...963...83H,
	author = {{Huang}, Caroline D. and {Yuan}, Wenlong and {Riess}, Adam G. and {Hack}, Warren and {Whitelock}, Patricia A. and {Zakamska}, Nadia L. and {Casertano}, Stefano and {Macri}, Lucas M. and {Marengo}, Massimo and {Menzies}, John W. and {Smith}, Randall K.},
	title = "{The Mira Distance to M101 and a 4\% Measurement of H $_{0}$}",
	journal = {\apj},
	year = 2024,
	month = mar,
	volume = {963},
	number = {2},
	eid = {83},
	pages = {83},
	doi = {10.3847/1538-4357/ad1ff8},
	archivePrefix = {arXiv},
	eprint = {2312.08423},
	primaryClass = {astro-ph.CO},
	adsurl = {https://ui.adsabs.harvard.edu/abs/2024ApJ...963...83H}
}

@ARTICLE{H0DN2026A&A...708A.166H,
	author = {{H0DN Collaboration} and {Casertano}, Stefano and {Anand}, Gagandeep and {Anderson}, Richard I. and {Beaton}, Rachael and {Bhardwaj}, Anupam and {Blakeslee}, John P. and {Boubel}, Paula and {Breuval}, Louise and {Brout}, Dillon and {Cantiello}, Michele and {Cruz Reyes}, Mauricio and {Cs{\"o}rnyei}, Geza and {de Jaeger}, Thomas and {Dhawan}, Suhail and {Di Valentino}, Eleonora and {Galbany}, Llu{\'\i}s and {Gil-Mar{\'\i}n}, H{\'e}ctor and {Graczyk}, Dariusz and {Huang}, Caroline and {Jensen}, Joseph B. and {Kervella}, Pierre and {Leibundgut}, Bruno and {Lengen}, Bastian and {Li}, Siyang and {Macri}, Lucas and {{\"O}z{\"u}lker}, Emre and {Pesce}, Dominic W. and {Riess}, Adam and {Romaniello}, Martino and {Said}, Khaled and {Sch{\"o}neberg}, Nils and {Scolnic}, Dan and {Sicignano}, Teresa and {Skowron}, Dorota M. and {Uddin}, Syed A. and {Verde}, Licia and {Nota}, Antonella},
	title = "{The Local Distance Network: A community consensus report on the measurement of the Hubble constant at {\ensuremath{\sim}}1\% precision}",
	journal = {\aap},
	year = 2026,
	month = apr,
	volume = {708},
	eid = {A166},
	pages = {A166},
	doi = {10.1051/0004-6361/202557993},
	archivePrefix = {arXiv},
	eprint = {2510.23823},
	primaryClass = {astro-ph.CO},
	adsurl = {https://ui.adsabs.harvard.edu/abs/2026A&A...708A.166H}
}

@ARTICLE{Chiavassa2018A&A...617L...1C,
	author = {{Chiavassa}, A. and {Freytag}, B. and {Schultheis}, M.},
	title = "{Heading Gaia to measure atmospheric dynamics in AGB stars}",
	journal = {\aap},
	year = 2018,
	month = sep,
	volume = {617},
	eid = {L1},
	pages = {L1},
	doi = {10.1051/0004-6361/201833844},
	archivePrefix = {arXiv},
	eprint = {1808.02548},
	primaryClass = {astro-ph.SR},
	adsurl = {https://ui.adsabs.harvard.edu/abs/2018A&A...617L...1C}
}

@ARTICLE{Huang2018ApJ...857...67H,
	author = {{Huang}, Caroline D. and {Riess}, Adam G. and {Hoffmann}, Samantha L. and {Klein}, Christopher and {Bloom}, Joshua and {Yuan}, Wenlong and {Macri}, Lucas M. and {Jones}, David O. and {Whitelock}, Patricia A. and {Casertano}, Stefano and {Anderson}, Richard I.},
	title = "{A Near-infrared Period-Luminosity Relation for Miras in NGC 4258, an Anchor for a New Distance Ladder}",
	journal = {\apj},
	year = 2018,
	month = apr,
	volume = {857},
	number = {1},
	eid = {67},
	pages = {67},
	doi = {10.3847/1538-4357/aab6b3},
	archivePrefix = {arXiv},
	eprint = {1801.02711},
	primaryClass = {astro-ph.CO},
	adsurl = {https://ui.adsabs.harvard.edu/abs/2018ApJ...857...67H}
}

@ARTICLE{Huang2020ApJ...889....5H,
	author = {{Huang}, Caroline D. and {Riess}, Adam G. and {Yuan}, Wenlong and {Macri}, Lucas M. and {Zakamska}, Nadia L. and {Casertano}, Stefano and {Whitelock}, Patricia A. and {Hoffmann}, Samantha L. and {Filippenko}, Alexei V. and {Scolnic}, Daniel},
	title = "{Hubble Space Telescope Observations of Mira Variables in the SN Ia Host NGC 1559: An Alternative Candle to Measure the Hubble Constant}",
	journal = {\apj},
	year = 2020,
	month = jan,
	volume = {889},
	number = {1},
	eid = {5},
	pages = {5},
	doi = {10.3847/1538-4357/ab5dbd},
	archivePrefix = {arXiv},
	eprint = {1908.10883},
	primaryClass = {astro-ph.CO},
	adsurl = {https://ui.adsabs.harvard.edu/abs/2020ApJ...889....5H}
}

@ARTICLE{Yuan2017AJ....154..149Y,
	author = {{Yuan}, Wenlong and {Macri}, Lucas M. and {He}, Shiyuan and {Huang}, Jianhua Z. and {Kanbur}, Shashi M. and {Ngeow}, Chow-Choong},
	title = "{Large Magellanic Cloud Near-infrared Synoptic Survey. V. Period-Luminosity Relations of Miras}",
	journal = {\aj},
	year = 2017,
	month = oct,
	volume = {154},
	number = {4},
	eid = {149},
	pages = {149},
	doi = {10.3847/1538-3881/aa86f1},
	archivePrefix = {arXiv},
	eprint = {1708.04742},
	primaryClass = {astro-ph.SR},
	adsurl = {https://ui.adsabs.harvard.edu/abs/2017AJ....154..149Y}
}

@ARTICLE{Whitelock2008MNRAS.386..313W,
	author = {{Whitelock}, Patricia A. and {Feast}, Michael W. and {Van Leeuwen}, Floor},
	title = "{AGB variables and the Mira period-luminosity relation}",
	journal = {\mnras},
	year = 2008,
	month = may,
	volume = {386},
	number = {1},
	pages = {313-323},
	doi = {10.1111/j.1365-2966.2008.13032.x},
	archivePrefix = {arXiv},
	eprint = {0801.4465},
	primaryClass = {astro-ph},
	adsurl = {https://ui.adsabs.harvard.edu/abs/2008MNRAS.386..313W}
}

@ARTICLE{Breuval2024ApJ...973...30B,
	author = {{Breuval}, Louise and {Riess}, Adam G. and {Casertano}, Stefano and {Yuan}, Wenlong and {Macri}, Lucas M. and {Romaniello}, Martino and {Murakami}, Yukei S. and {Scolnic}, Daniel and {Anand}, Gagandeep S. and {Soszy{\'n}ski}, Igor},
	title = "{Small Magellanic Cloud Cepheids Observed with the Hubble Space Telescope Provide a New Anchor for the SH0ES Distance Ladder}",
	journal = {\apj},
	year = 2024,
	month = sep,
	volume = {973},
	number = {1},
	eid = {30},
	pages = {30},
	doi = {10.3847/1538-4357/ad630e},
	archivePrefix = {arXiv},
	eprint = {2404.08038},
	primaryClass = {astro-ph.CO},
	adsurl = {https://ui.adsabs.harvard.edu/abs/2024ApJ...973...30B}
}

@ARTICLE{Riess2021ApJ...908L...6R,
	author = {{Riess}, Adam G. and {Casertano}, Stefano and {Yuan}, Wenlong and {Bowers}, J. Bradley and {Macri}, Lucas and {Zinn}, Joel C. and {Scolnic}, Dan},
	title = "{Cosmic Distances Calibrated to 1\% Precision with Gaia EDR3 Parallaxes and Hubble Space Telescope Photometry of 75 Milky Way Cepheids Confirm Tension with {\ensuremath{\Lambda}}CDM}",
	journal = {\apjl},
	year = 2021,
	month = feb,
	volume = {908},
	number = {1},
	eid = {L6},
	pages = {L6},
	doi = {10.3847/2041-8213/abdbaf},
	archivePrefix = {arXiv},
	eprint = {2012.08534},
	primaryClass = {astro-ph.CO},
	adsurl = {https://ui.adsabs.harvard.edu/abs/2021ApJ...908L...6R}
}

@ARTICLE{Kervella2019A&A...623A.116K,
	author = {{Kervella}, Pierre and {Gallenne}, Alexandre and {Remage Evans}, Nancy and {Szabados}, Laszlo and {Arenou}, Fr{\'e}d{\'e}ric and {M{\'e}rand}, Antoine and {Proto}, Yann and {Karczmarek}, Paulina and {Nardetto}, Nicolas and {Gieren}, Wolfgang and {Pietrzynski}, Grzegorz},
	title = "{Multiplicity of Galactic Cepheids and RR Lyrae stars from Gaia DR2. I. Binarity from proper motion anomaly}",
	journal = {\aap},
	year = 2019,
	month = mar,
	volume = {623},
	eid = {A116},
	pages = {A116},
	doi = {10.1051/0004-6361/201834210},
	archivePrefix = {arXiv},
	eprint = {1903.03632},
	primaryClass = {astro-ph.SR},
	adsurl = {https://ui.adsabs.harvard.edu/abs/2019A&A...623A.116K}
}

@ARTICLE{Evans2015AJ....150...13E,
	author = {{Evans}, Nancy Remage and {Berdnikov}, Leonid and {Lauer}, Jennifer and {Morgan}, Douglas and {Nichols}, Joy and {G{\"u}nther}, H. Moritz and {Gorynya}, Natalya and {Rastorguev}, Alexey and {Moskalik}, Pawel},
	title = "{Binary Properties from Cepheid Radial Velocities (CRaV)}",
	journal = {\aj},
	year = 2015,
	month = jul,
	volume = {150},
	number = {1},
	eid = {13},
	pages = {13},
	doi = {10.1088/0004-6256/150/1/13},
	archivePrefix = {arXiv},
	eprint = {1505.05823},
	primaryClass = {astro-ph.SR},
	adsurl = {https://ui.adsabs.harvard.edu/abs/2015AJ....150...13E}
}

@ARTICLE{Halbwachs_2023,
       author = {{Halbwachs}, Jean-Louis and {Pourbaix}, Dimitri and {Arenou}, Fr{\'e}d{\'e}ric and {Galluccio}, Laurent and {Guillout}, Patrick and {Bauchet}, Nathalie and {Marchal}, Olivier and {Sadowski}, Gilles and {Teyssier}, David},
        title = "{Gaia Data Release 3. Astrometric binary star processing}",
      journal = {\aap},
         year = 2023,
        month = jun,
       volume = {674},
          eid = {A9},
        pages = {A9},
          doi = {10.1051/0004-6361/202243969},
archivePrefix = {arXiv},
       eprint = {2206.05726},
 primaryClass = {astro-ph.SR},
       adsurl = {https://ui.adsabs.harvard.edu/abs/2023A&A...674A...9H}
}

@ARTICLE{DR3,
       author = {{Gaia Collaboration} and {Vallenari}, A. and {Brown}, A.~G.~A. and {Prusti}, T. and {de Bruijne}, J.~H.~J. and {Arenou}, F. and {Babusiaux}, C. and {Biermann}, M. and {Creevey}, O.~L. and {Ducourant}, C. and et al.},
        title = "{Gaia Data Release 3. Summary of the content and survey properties}",
      journal = {\aap},
         year = 2023,
        month = jun,
       volume = {674},
          eid = {A1},
        pages = {A1},
          doi = {10.1051/0004-6361/202243940},
archivePrefix = {arXiv},
       eprint = {2208.00211},
 primaryClass = {astro-ph.GA},
       adsurl = {https://ui.adsabs.harvard.edu/abs/2023A&A...674A...1G}
}

@ARTICLE{Gaia,
       author = {{Gaia Collaboration} and {Prusti}, T. and {de Bruijne}, J.~H.~J. and {Brown}, A.~G.~A. and {Vallenari}, A. and {Babusiaux}, C. and {Bailer-Jones}, C.~A.~L. and {Bastian}, U. and {Biermann}, M. and {Evans}, D.~W. and et al.},
        title = "{The Gaia mission}",
      journal = {\aap},
         year = 2016,
        month = nov,
       volume = {595},
          eid = {A1},
        pages = {A1},
          doi = {10.1051/0004-6361/201629272},
archivePrefix = {arXiv},
       eprint = {1609.04153},
 primaryClass = {astro-ph.IM},
       adsurl = {https://ui.adsabs.harvard.edu/abs/2016A&A...595A...1G}
}

@ARTICLE{Wielen_1996,
       author = {{Wielen}, R.},
        title = "{Searching for VIMs: an astrometric method to detect the binary nature of double stars with a variable component}",
      journal = {\aap},
         year = 1996,
        month = oct,
       volume = {314},
        pages = {679},
       adsurl = {https://ui.adsabs.harvard.edu/abs/1996A&A...314..679W}
}

@ARTICLE{Kervella_2022,
       author = {{Kervella}, Pierre and {Arenou}, Fr{\'e}d{\'e}ric and {Th{\'e}venin}, Fr{\'e}d{\'e}ric},
        title = "{Stellar and substellar companions from Gaia EDR3. Proper-motion anomaly and resolved common proper-motion pairs}",
      journal = {\aap},
         year = 2022,
        month = jan,
       volume = {657},
          eid = {A7},
        pages = {A7},
          doi = {10.1051/0004-6361/202142146},
archivePrefix = {arXiv},
       eprint = {2109.10912},
 primaryClass = {astro-ph.SR},
       adsurl = {https://ui.adsabs.harvard.edu/abs/2022A&A...657A...7K}
}

@ARTICLE{Decin2020Sci...369.1497D,
	author = {{Decin}, L. and {Montarg{\`e}s}, M. and {Richards}, A.~M.~S. and {Gottlieb}, C.~A. and {Homan}, W. and {McDonald}, I. and {El Mellah}, I. and {Danilovich}, T. and {Wallstr{\"o}m}, S.~H.~J. and {Zijlstra}, A. and {Baudry}, A. and {Bolte}, J. and {Cannon}, E. and {De Beck}, E. and {De Ceuster}, F. and {de Koter}, A. and {De Ridder}, J. and {Etoka}, S. and {Gobrecht}, D. and {Gray}, M. and {Herpin}, F. and {Jeste}, M. and {Lagadec}, E. and {Kervella}, P. and {Khouri}, T. and {Menten}, K. and {Millar}, T.~J. and {M{\"u}ller}, H.~S.~P. and {Plane}, J.~M.~C. and {Sahai}, R. and {Sana}, H. and {Van de Sande}, M. and {Waters}, L.~B.~F.~M. and {Wong}, K.~T. and {Yates}, J.},
	title = "{(Sub)stellar companions shape the winds of evolved stars}",
	journal = {Science},
	year = 2020,
	month = sep,
	volume = {369},
	number = {6510},
	pages = {1497-1500},
	doi = {10.1126/science.abb1229},
	archivePrefix = {arXiv},
	eprint = {2009.11694},
	primaryClass = {astro-ph.SR},
	adsurl = {https://ui.adsabs.harvard.edu/abs/2020Sci...369.1497D}
}

@ARTICLE{Moe2017ApJS..230...15M,
       author = {{Moe}, Maxwell and {Di Stefano}, Rosanne},
        title = "{Mind Your Ps and Qs: The Interrelation between Period (P) and Mass-ratio (Q) Distributions of Binary Stars}",
      journal = {\apjs},
         year = 2017,
        month = jun,
       volume = {230},
       number = {2},
          eid = {15},
        pages = {15},
          doi = {10.3847/1538-4365/aa6fb6},
archivePrefix = {arXiv},
       eprint = {1606.05347},
 primaryClass = {astro-ph.SR},
       adsurl = {https://ui.adsabs.harvard.edu/abs/2017ApJS..230...15M}
}

@ARTICLE{Fulton2018AJ....156..264F,
       author = {{Fulton}, Benjamin J. and {Petigura}, Erik A.},
        title = "{The California-Kepler Survey. VII. Precise Planet Radii Leveraging Gaia DR2 Reveal the Stellar Mass Dependence of the Planet Radius Gap}",
      journal = {\aj},
         year = 2018,
        month = dec,
       volume = {156},
       number = {6},
          eid = {264},
        pages = {264},
          doi = {10.3847/1538-3881/aae828},
archivePrefix = {arXiv},
       eprint = {1805.01453},
 primaryClass = {astro-ph.EP},
       adsurl = {https://ui.adsabs.harvard.edu/abs/2018AJ....156..264F}
}

@ARTICLE{Watson2006SASS...25...47W,
       author = {{Watson}, C.~L. and {Henden}, A.~A. and {Price}, A.},
        title = "{The International Variable Star Index (VSX)}",
      journal = {Society for Astronomical Sciences Annual Symposium},
         year = 2006,
        month = may,
       volume = {25},
        pages = {47},
       adsurl = {https://ui.adsabs.harvard.edu/abs/2006SASS...25...47W}
}

@ARTICLE{Paladini2018Natur.553..310P,
	author = {{Paladini}, C. and {Baron}, F. and {Jorissen}, A. and {Le Bouquin}, J.-B. and {Freytag}, B. and {van Eck}, S. and {Wittkowski}, M. and {Hron}, J. and {Chiavassa}, A. and {Berger}, J.-P. and {Siopis}, C. and {Mayer}, A. and {Sadowski}, G. and {Kravchenko}, K. and {Shetye}, S. and {Kerschbaum}, F. and {Kluska}, J. and {Ramstedt}, S.},
	title = "{Large granulation cells on the surface of the giant star {\ensuremath{\pi}}$^{1}$ Gruis}",
	journal = {\nat},
	year = 2018,
	month = jan,
	volume = {553},
	number = {7688},
	pages = {310-312},
	doi = {10.1038/nature25001},
	adsurl = {https://ui.adsabs.harvard.edu/abs/2018Natur.553..310P}
}

@ARTICLE{Vlemmings2024Natur.633..323V,
       author = {{Vlemmings}, Wouter and {Khouri}, Theo and {Bojnordi Arbab}, Behzad and {De Beck}, Elvire and {Maercker}, Matthias},
        title = "{One month convection timescale on the surface of a giant evolved star}",
      journal = {\nat},
         year = 2024,
        month = sep,
       volume = {633},
       number = {8029},
        pages = {323-326},
          doi = {10.1038/s41586-024-07836-9},
archivePrefix = {arXiv},
       eprint = {2409.06785},
 primaryClass = {astro-ph.SR},
       adsurl = {https://ui.adsabs.harvard.edu/abs/2024Natur.633..323V}
}

@ARTICLE{Montarges2025A&A...699A..22M,
       author = {{Montarg{\`e}s}, M. and {Malfait}, J. and {Esseldeurs}, M. and {de Koter}, A. and {Baron}, F. and {Kervella}, P. and {Danilovich}, T. and {Richards}, A.~M.~S. and {Sahai}, R. and {McDonald}, I. and {Khouri}, T. and {Shetye}, S. and {Zijlstra}, A. and {Van de Sande}, M. and {El Mellah}, I. and {Herpin}, F. and {Siess}, L. and {Etoka}, S. and {Gobrecht}, D. and {Marinho}, L. and {Wallstr{\"o}m}, S.~H.~J. and {Wong}, K.~T. and {Yates}, J.},
        title = "{An accreting dwarf star orbiting the S-type giant star {\ensuremath{\pi}}$^{1}$ Gru}",
      journal = {\aap},
         year = 2025,
        month = jul,
       volume = {699},
          eid = {A22},
        pages = {A22},
          doi = {10.1051/0004-6361/202452587},
archivePrefix = {arXiv},
       eprint = {2504.16845},
 primaryClass = {astro-ph.SR},
       adsurl = {https://ui.adsabs.harvard.edu/abs/2025A&A...699A..22M}
}

@ARTICLE{Esseldeurs2026NatAs..10..124E,
       author = {{Esseldeurs}, Mats and {Decin}, Leen and {De Ridder}, Joris and {Mori}, Yoshiya and {Karakas}, Amanda I. and {Malfait}, Jolien and {Danilovich}, Ta{\'\i}ssa and {Mathis}, St{\'e}phane and {Richards}, Anita M.~S. and {Sahai}, Raghvendra and {Yates}, Jeremy and {Van de Sande}, Marie and {Baes}, Maarten and {Baudry}, Alain and {Bolte}, Jan and {Ceulemans}, Thomas and {De Ceuster}, Frederik and {El Mellah}, Ileyk and {Etoka}, Sandra and {Gottlieb}, Carl and {Herpin}, Fabrice and {Kervella}, Pierre and {Landri}, Camille and {Marinho}, Louise and {McDonald}, Iain and {Menten}, Karl and {Millar}, Tom and {Osborn}, Zara and {Pimpanuwat}, Bannawit and {Plane}, John and {Price}, Daniel J. and {Siess}, Lionel and {Vermeulen}, Owen and {Wong}, Ka Tat},
        title = "{Evidence for the Keplerian orbit of a close companion around a giant star}",
      journal = {Nature Astronomy},
         year = 2026,
        month = jan,
       volume = {10},
        pages = {124-143},
          doi = {10.1038/s41550-025-02697-2},
archivePrefix = {arXiv},
       eprint = {2511.11247},
 primaryClass = {astro-ph.SR},
       adsurl = {https://ui.adsabs.harvard.edu/abs/2026NatAs..10..124E}
}

@INPROCEEDINGS{Offner2023ASPC..534..275O,
       author = {{Offner}, S.~S.~R. and {Moe}, M. and {Kratter}, K.~M. and {Sadavoy}, S.~I. and {Jensen}, E.~L.~N. and {Tobin}, J.~J.},
        title = "{The Origin and Evolution of Multiple Star Systems}",
    booktitle = {Protostars and Planets VII},
         year = 2023,
       editor = {{Inutsuka}, S. and {Aikawa}, Y. and {Muto}, T. and {Tomida}, K. and {Tamura}, M.},
       series = {Astronomical Society of the Pacific Conference Series},
       volume = {534},
        month = jul,
        pages = {275},
          doi = {10.48550/arXiv.2203.10066},
archivePrefix = {arXiv},
       eprint = {2203.10066},
 primaryClass = {astro-ph.SR},
       adsurl = {https://ui.adsabs.harvard.edu/abs/2023ASPC..534..275O}
}

@ARTICLE{Lindegren2021A&A...649A...4L,
	author = {{Lindegren}, L. and {Bastian}, U. and {Biermann}, M. and {Bombrun}, A. and {de Torres}, A. and {Gerlach}, E. and {Geyer}, R. and {Hern{\'a}ndez}, J. and {Hilger}, T. and {Hobbs}, D. and {Klioner}, S.~A. and {Lammers}, U. and {McMillan}, P.~J. and {Ramos-Lerate}, M. and {Steidelm{\"u}ller}, H. and {Stephenson}, C.~A. and {van Leeuwen}, F.},
	title = "{Gaia Early Data Release 3. Parallax bias versus magnitude, colour, and position}",
	journal = {\aap},
	year = 2021,
	month = may,
	volume = {649},
	eid = {A4},
	pages = {A4},
	doi = {10.1051/0004-6361/202039653},
	archivePrefix = {arXiv},
	eprint = {2012.01742},
	primaryClass = {astro-ph.IM},
	adsurl = {https://ui.adsabs.harvard.edu/abs/2021A&A...649A...4L}
}

@ARTICLE{Lindegren2021A&A...649A...2L,
       author = {{Lindegren}, L. and {Klioner}, S.~A. and {Hern{\'a}ndez}, J. and {Bombrun}, A. and {Ramos-Lerate}, M. and {Steidelm{\"u}ller}, H. and {Bastian}, U. and {Biermann}, M. and {de Torres}, A. and {Gerlach}, E. and {Geyer}, R. and {Hilger}, T. and {Hobbs}, D. and {Lammers}, U. and {McMillan}, P.~J. and {Stephenson}, C.~A. and {Casta{\~n}eda}, J. and {Davidson}, M. and {Fabricius}, C. and {Gracia-Abril}, G. and {Portell}, J. and {Rowell}, N. and {Teyssier}, D. and {Torra}, F. and {Bartolom{\'e}}, S. and {Clotet}, M. and {Garralda}, N. and {Gonz{\'a}lez-Vidal}, J.~J. and {Torra}, J. and {Abbas}, U. and {Altmann}, M. and {Anglada Varela}, E. and {Balaguer-N{\'u}{\~n}ez}, L. and {Balog}, Z. and {Barache}, C. and {Becciani}, U. and {Bernet}, M. and {Bertone}, S. and {Bianchi}, L. and {Bouquillon}, S. and {Brown}, A.~G.~A. and {Bucciarelli}, B. and {Busonero}, D. and {Butkevich}, A.~G. and {Buzzi}, R. and {Cancelliere}, R. and {Carlucci}, T. and {Charlot}, P. and {Cioni}, M.-R.~L. and {Crosta}, M. and {Crowley}, C. and {del Peloso}, E.~F. and {del Pozo}, E. and {Drimmel}, R. and {Esquej}, P. and {Fienga}, A. and {Fraile}, E. and {Gai}, M. and {Garcia-Reinaldos}, M. and {Guerra}, R. and {Hambly}, N.~C. and {Hauser}, M. and {Jan{\ss}en}, K. and {Jordan}, S. and {Kostrzewa-Rutkowska}, Z. and {Lattanzi}, M.~G. and {Liao}, S. and {Licata}, E. and {Lister}, T.~A. and {L{\"o}ffler}, W. and {Marchant}, J.~M. and {Masip}, A. and {Mignard}, F. and {Mints}, A. and {Molina}, D. and {Mora}, A. and {Morbidelli}, R. and {Murphy}, C.~P. and {Pagani}, C. and {Panuzzo}, P. and {Pe{\~n}alosa Esteller}, X. and {Poggio}, E. and {Re Fiorentin}, P. and {Riva}, A. and {Sagrist{\`a} Sell{\'e}s}, A. and {Sanchez Gimenez}, V. and {Sarasso}, M. and {Sciacca}, E. and {Siddiqui}, H.~I. and {Smart}, R.~L. and {Souami}, D. and {Spagna}, A. and {Steele}, I.~A. and {Taris}, F. and {Utrilla}, E. and {van Reeven}, W. and {Vecchiato}, A.},
        title = "{Gaia Early Data Release 3. The astrometric solution}",
      journal = {\aap},
         year = 2021,
        month = may,
       volume = {649},
          eid = {A2},
        pages = {A2},
          doi = {10.1051/0004-6361/202039709},
archivePrefix = {arXiv},
       eprint = {2012.03380},
 primaryClass = {astro-ph.IM},
       adsurl = {https://ui.adsabs.harvard.edu/abs/2021A&A...649A...2L}
}

@ARTICLE{Chiavassa2020A&A...640A..23C,
	author = {{Chiavassa}, A. and {Kravchenko}, K. and {Millour}, F. and {Schaefer}, G. and {Schultheis}, M. and {Freytag}, B. and {Creevey}, O. and {Hocd{\'e}}, V. and {Morand}, F. and {Ligi}, R. and {Kraus}, S. and {Monnier}, J.~D. and {Mourard}, D. and {Nardetto}, N. and {Anugu}, N. and {Le Bouquin}, J.-B. and {Davies}, C.~L. and {Ennis}, J. and {Gardner}, T. and {Labdon}, A. and {Lanthermann}, C. and {Setterholm}, B.~R. and {ten Brummelaar}, T.},
	title = "{Optical interferometry and Gaia measurement uncertainties reveal the physics of asymptotic giant branch stars}",
	journal = {\aap},
	year = 2020,
	month = aug,
	volume = {640},
	eid = {A23},
	pages = {A23},
	doi = {10.1051/0004-6361/202037832},
	archivePrefix = {arXiv},
	eprint = {2006.07318},
	primaryClass = {astro-ph.SR},
	adsurl = {https://ui.adsabs.harvard.edu/abs/2020A&A...640A..23C}
}

@ARTICLE{Wood2000PASA...17...18W,
	author = {{Wood}, P.~R.},
	title = "{Variable Red Giants in the LMC: Pulsating Stars and Binaries?}",
	journal = {\pasa},
	year = 2000,
	month = apr,
	volume = {17},
	number = {1},
	pages = {18-21},
	doi = {10.1071/AS00018},
	adsurl = {https://ui.adsabs.harvard.edu/abs/2000PASA...17...18W}
}

@ARTICLE{Trabucchi2021A&A...656A..66T,
	author = {{Trabucchi}, M. and {Mowlavi}, N. and {Lebzelter}, T.},
	title = "{Semi-regular red giants as distance indicators. I. The period-luminosity relations of semi-regular variables revisited}",
	journal = {\aap},
	year = 2021,
	month = dec,
	volume = {656},
	eid = {A66},
	pages = {A66},
	doi = {10.1051/0004-6361/202142022},
	archivePrefix = {arXiv},
	eprint = {2109.04293},
	primaryClass = {astro-ph.SR},
	adsurl = {https://ui.adsabs.harvard.edu/abs/2021A&A...656A..66T}
}

@INCOLLECTION{Jorissen2004agbs.book..461J,
       author = {{Jorissen}, Alain},
        title = "{AGB Stars in Binaries and Their Progeny}",
    booktitle = {Asymptotic Giant Branch Stars},
         year = 2004,
       editor = {{Habing}, Harm J. and {Olofsson}, Hans},
        pages = {461-518},
          doi = {10.1007/978-1-4757-3876-6_9},
          publisher = {Springer-Verlag New York, Inc.},
       adsurl = {https://ui.adsabs.harvard.edu/abs/2004agbs.book..461J}
}

@ARTICLE{Planquart2024A&A...682A.143P,
       author = {{Planquart}, L. and {Jorissen}, A. and {Escorza}, A. and {Verhamme}, O. and {Van Winckel}, H.},
        title = "{A dynamic view of V Hydrae. Monitoring of a spectroscopic-binary AGB star with an alkaline jet}",
      journal = {\aap},
         year = 2024,
        month = feb,
       volume = {682},
          eid = {A143},
        pages = {A143},
          doi = {10.1051/0004-6361/202347947},
archivePrefix = {arXiv},
       eprint = {2405.07820},
 primaryClass = {astro-ph.SR},
       adsurl = {https://ui.adsabs.harvard.edu/abs/2024A&A...682A.143P}
}

@ARTICLE{Gaia2023A&A...674A..34G,
	author = {{Gaia Collaboration} and {Arenou}, F. and {Babusiaux}, C. and {Barstow}, M.~A. and {Faigler}, S. and {Jorissen}, A. and {Kervella}, P. and {Mazeh}, T. and {Mowlavi}, N. and {Panuzzo}, P. and {Sahlmann}, J. and {Shahaf}, S. and {Sozzetti}, A. and {Bauchet}, N. and {Damerdji}, Y. and {Gavras}, P. and {Giacobbe}, P. and {Gosset}, E. and {Halbwachs}, J.-L. and {Holl}, B. and {Lattanzi}, M.~G. and {Leclerc}, N. and {Morel}, T. and {Pourbaix}, D. and {Re Fiorentin}, P. and {Sadowski}, G. and {S{\'e}gransan}, D. and {Siopis}, C. and {Teyssier}, D. and {Zwitter}, T. and {Planquart}, L. and {Brown}, A.~G.~A. and {Vallenari}, A. and {Prusti}, T. and {de Bruijne}, J.~H.~J. and {Biermann}, M. and {Creevey}, O.~L. and {Ducourant}, C. and {Evans}, D.~W. and {Eyer}, L. and {Guerra}, R. and {Hutton}, A. and {Jordi}, C. and {Klioner}, S.~A. and {Lammers}, U.~L. and {Lindegren}, L. and {Luri}, X. and {Mignard}, F. and {Panem}, C. and {Randich}, S. and {Sartoretti}, P. and {Soubiran}, C. and {Tanga}, P. and {Walton}, N.~A. and {Bailer-Jones}, C.~A.~L. and {Bastian}, U. and {Drimmel}, R. and {Jansen}, F. and {Katz}, D. and {van Leeuwen}, F. and {Bakker}, J. and {Cacciari}, C. and {Casta{\~n}eda}, J. and {De Angeli}, F. and {Fabricius}, C. and {Fouesneau}, M. and {Fr{\'e}mat}, Y. and {Galluccio}, L. and {Guerrier}, A. and {Heiter}, U. and {Masana}, E. and {Messineo}, R. and {Nicolas}, C. and {Nienartowicz}, K. and {Pailler}, F. and {Riclet}, F. and {Roux}, W. and {Seabroke}, G.~M. and {Sordo}, R. and {Th{\'e}venin}, F. and {Gracia-Abril}, G. and {Portell}, J. and {Altmann}, M. and {Andrae}, R. and {Audard}, M. and {Bellas-Velidis}, I. and {Benson}, K. and {Berthier}, J. and {Blomme}, R. and {Burgess}, P.~W. and {Busonero}, D. and {Busso}, G. and {C{\'a}novas}, H. and {Carry}, B. and {Cellino}, A. and {Cheek}, N. and {Clementini}, G. and {Davidson}, M. and {de Teodoro}, P. and {Nu{\~n}ez Campos}, M. and {Delchambre}, L. and {Dell'Oro}, A. and {Esquej}, P. and {Fern{\'a}ndez-Hern{\'a}ndez}, J. and {Fraile}, E. and {Garabato}, D. and {Garc{\'\i}a-Lario}, P. and {Haigron}, R. and {Hambly}, N.~C. and {Harrison}, D.~L. and {Hern{\'a}ndez}, J. and {Hestroffer}, D. and {Hodgkin}, S.~T. and {Jan{\ss}en}, K. and {Jevardat de Fombelle}, G. and {Jordan}, S. and {Krone-Martins}, A. and {Lanzafame}, A.~C. and {L{\"o}ffler}, W. and {Marchal}, O. and {Marrese}, P.~M. and {Moitinho}, A. and {Muinonen}, K. and {Osborne}, P. and {Pancino}, E. and {Pauwels}, T. and {Recio-Blanco}, A. and {Reyl{\'e}}, C. and {Riello}, M. and {Rimoldini}, L. and {Roegiers}, T. and {Rybizki}, J. and {Sarro}, L.~M. and {Smith}, M. and {Utrilla}, E. and {van Leeuwen}, M. and {Abbas}, U. and {{\'A}brah{\'a}m}, P. and {Abreu Aramburu}, A. and {Aerts}, C. and {Aguado}, J.~J. and {Ajaj}, M. and {Aldea-Montero}, F. and {Altavilla}, G. and {{\'A}lvarez}, M.~A. and {Alves}, J. and {Anders}, F. and {Anderson}, R.~I. and {Anglada Varela}, E. and {Antoja}, T. and {Baines}, D. and {Baker}, S.~G. and {Balaguer-N{\'u}{\~n}ez}, L. and {Balbinot}, E. and {Balog}, Z. and {Barache}, C. and {Barbato}, D. and {Barros}, M. and {Bartolom{\'e}}, S. and {Bassilana}, J.-L. and {Becciani}, U. and {Bellazzini}, M. and {Berihuete}, A. and {Bernet}, M. and {Bertone}, S. and {Bianchi}, L. and {Binnenfeld}, A. and {Blanco-Cuaresma}, S. and {Blazere}, A. and {Boch}, T. and {Bombrun}, A. and {Bossini}, D. and {Bouquillon}, S. and {Bragaglia}, A. and {Bramante}, L. and {Breedt}, E. and {Bressan}, A. and {Brouillet}, N. and {Brugaletta}, E. and {Bucciarelli}, B. and {Burlacu}, A. and {Butkevich}, A.~G. and {Buzzi}, R. and {Caffau}, E. and {Cancelliere}, R. and {Cantat-Gaudin}, T. and {Carballo}, R. and {Carlucci}, T. and {Carnerero}, M.~I. and {Carrasco}, J.~M. and {Casamiquela}, L. and {Castellani}, M. and {Castro-Ginard}, A. and {Chaoul}, L. and {Charlot}, P. and {Chemin}, L. and {Chiaramida}, V. and {Chiavassa}, A. and {Chornay}, N. and {Comoretto}, G.},
	title = "{Gaia Data Release 3. Stellar multiplicity, a teaser for the hidden treasure}",
	journal = {\aap},
	year = 2023,
	month = jun,
	volume = {674},
	eid = {A34},
	pages = {A34},
	doi = {10.1051/0004-6361/202243782},
	archivePrefix = {arXiv},
	eprint = {2206.05595},
	primaryClass = {astro-ph.SR},
	adsurl = {https://ui.adsabs.harvard.edu/abs/2023A&A...674A..34G}
}

@ARTICLE{Beguin_2024,
       author = {{B{\'e}guin}, E. and {Chiavassa}, A. and {Ahmad}, A. and {Freytag}, B. and {Uttenthaler}, S.},
        title = "{Retrieving stellar parameters and dynamics of AGB stars with Gaia parallax measurements and CO$^{5}$BOLD RHD simulations}",
      journal = {\aap},
         year = 2024,
        month = oct,
       volume = {690},
          eid = {A125},
        pages = {A125},
          doi = {10.1051/0004-6361/202450245},
archivePrefix = {arXiv},
       eprint = {2409.03422},
 primaryClass = {astro-ph.SR},
       adsurl = {https://ui.adsabs.harvard.edu/abs/2024A&A...690A.125B}
}

@ARTICLE{Lindegren_2012,
       author = {{Lindegren}, L. and {Lammers}, U. and {Hobbs}, D. and {O'Mullane}, W. and {Bastian}, U. and {Hern{\'a}ndez}, J.},
        title = "{The astrometric core solution for the Gaia mission. Overview of models, algorithms, and software implementation}",
      journal = {\aap},
         year = 2012,
        month = feb,
       volume = {538},
          eid = {A78},
        pages = {A78},
          doi = {10.1051/0004-6361/201117905},
archivePrefix = {arXiv},
       eprint = {1112.4139},
 primaryClass = {astro-ph.IM},
       adsurl = {https://ui.adsabs.harvard.edu/abs/2012A&A...538A..78L}
}

@ARTICLE{ElBadry2024OJAp....7E.100E,
       author = {{El-Badry}, Kareem and {Lam}, Casey and {Holl}, Berry and {Halbwachs}, Jean-Louis and {Rix}, Hans-Walter and {Mazeh}, Tsevi and {Shahaf}, Sahar},
        title = "{A generative model for Gaia astrometric orbit catalogs: selection functions for binary stars, giant planets, and compact object companions}",
      journal = {The Open Journal of Astrophysics},
         year = 2024,
        month = nov,
       volume = {7},
          eid = {100},
        pages = {100},
          doi = {10.33232/001c.125461},
archivePrefix = {arXiv},
       eprint = {2411.00088},
 primaryClass = {astro-ph.SR},
       adsurl = {https://ui.adsabs.harvard.edu/abs/2024OJAp....7E.100E}
}

@ARTICLE{1953MNRAS.113..510F,
       author = {{Feast}, M.~W.},
        title = "{The absolute magnitude and spectrum of the class S star {\ensuremath{\pi}}$^{1}$ Gruis}",
      journal = {\mnras},
         year = 1953,
        month = jan,
       volume = {113},
        pages = {510},
          doi = {10.1093/mnras/113.4.510},
       adsurl = {https://ui.adsabs.harvard.edu/abs/1953MNRAS.113..510F}
}

\newpage
\onecolumn
\begin{appendix}
	\section{Additional figures and tables}

	\begin{figure*}[htp]
		\centering
		\includegraphics[width=\textwidth]{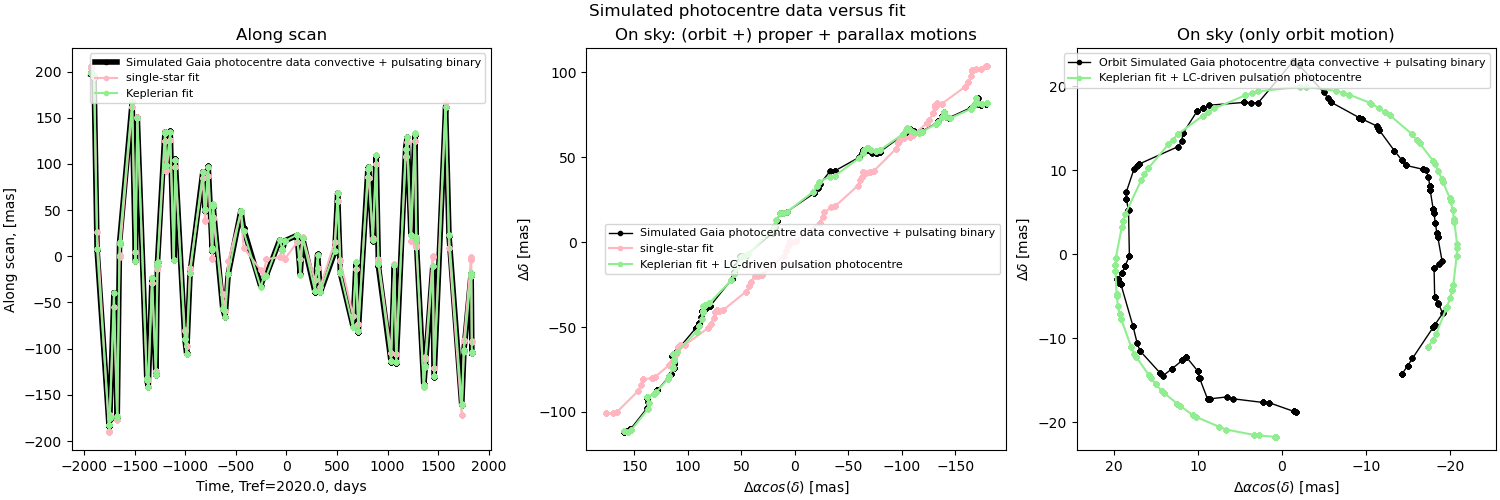}
		\caption{Same as Fig.~\ref{fig:pi1gru_retrieved_known_Vcomp}, but here for the case where convection-induced photocentre shifts were taken directly from the three-dimensional radiation-hydrodynamic model
			\texttt{st29gm04n001} of \citet{Beguin_2024} and the companion flux being unknown. Retrieved model parameters are listed in  Table~\ref{tab:pi1gru_hydro_rednoise}.}
		\label{fig:pi1gru_retrieved_unknown_Vcomp_conv_Chiav}
	\end{figure*}

\begin{table*}[htp]
	\centering
	\caption{Retrieved parameters for the single-star and VIM models fitted to the
		simulated $\pi^1$~Gru Gaia DR5 epoch astrometry without including red
		noise in the retrieval.
	}
	\label{tab:pi1gru_vim_no_rednoise}
	\begin{tabular}{lcccc}
		\hline\hline
		Parameter & SS & VIMF & VIML & VIMA \\
		\hline
		$\chi^2_{\rm red}$ & 467 & 408 & 398 & 136 \\
		\hline
		$\Delta \alpha^\star$ &
		$3.040 \pm 0.341$ &
		$4.215 \pm 0.338$ &
		$4.098 \pm 0.335$ &
		$6.184 \pm 0.306$ \\
		
		$\Delta \delta$ &
		$0.274 \pm 0.317$ &
		$0.146 \pm 0.301$ &
		$0.119 \pm 0.299$ &
		$13.383 \pm 0.297$ \\
		
		$\mu_{\alpha^\star}$ &
		$34.792 \pm 0.101$ &
		$34.769 \pm 0.095$ &
		$34.910 \pm 0.102$ &
		$35.362 \pm 0.070$ \\
		
		$\mu_\delta$ &
		$-21.517 \pm 0.100$ &
		$-21.351 \pm 0.094$ &
		$-21.549 \pm 0.100$ &
		$-21.200 \pm 0.060$ \\
		
		$\varpi$ &
		$6.187 \pm 0.458$ &
		$6.655 \pm 0.431$ &
		$6.417 \pm 0.429$ &
		$5.527 \pm 0.251$ \\
		
		\hline
		
		$D_\alpha$ &
		-- &
		$-7.379 \pm 1.979$ &
		$-7.426 \pm 1.962$ &
		$-17.306 \pm 1.247$ \\
		
		$D_\delta$ &
		-- &
		$23.858 \pm 1.779$ &
		$23.948 \pm 1.778$ &
		$-2.997 \pm 0.343$ \\
		
		$\dot{D}_\alpha$ &
		-- &
		-- &
		$-0.010 \pm 0.002$ &
		$-3.454 \pm 0.400$ \\
		
		$\dot{D}_\delta$ &
		-- &
		-- &
		$-0.004 \pm 0.002$ &
		$-0.733 \pm 0.356$ \\
		
		$k$ &
		-- &
		-- &
		-- &
		$-0.055 \pm 0.011$ \\
		
		$s$ &
		-- &
		-- &
		-- &
		$-2.7940 \pm 0.511$ \\
		
		\hline
		
		$\ddot{D}_\alpha$ &
		-- &
		-- &
		-- &
		$-0.166\pm 0.0003$ \\
		
		$\ddot{D}_\delta$ &
		-- &
		-- &
		-- &
		$0.959 \pm 0.015$ \\
		
		$\Delta\mu_{\alpha^\star}$ &
		-- &
		-- &
		-- &
		$-0.464 \pm 0.004$ \\
		
		$\Delta\mu_\delta$ &
		-- &
		-- &
		-- &
		$-2.680 \pm 0.004$ \\
		
		\hline
	\end{tabular}
\tablefoot{The first row lists the reduced chi-square values used in the uncertainty
	rescaling.}
\end{table*}

\begin{table*}[htp]
	\centering
	\caption{See caption of Table~\ref{tab:pi1gru_vim_no_rednoise}, but now listing the retrieved parameters for the simulated V~Hya Gaia DR5 epoch astrometry.}
	\label{tab:vhya_vim_no_rednoise}
	\begin{tabular}{lcccc}
		\hline\hline
		Parameter & SS & VIMF & VIML & VIMA \\
		\hline
		$\chi^2_{\rm red}$ & 525 & 521 & 506 & 89 \\
		\hline
		$\Delta \alpha^\star$ &
		$-2.419 \pm 0.142$ &
		$-2.786 \pm 0.166$ &
		$-2.633 \pm 0.165$ &
		$-5.471 \pm 0.094$ \\
		
		$\Delta \delta$ &
		$6.788 \pm 0.122$ &
		$6.791 \pm 0.134$ &
		$6.806 \pm 0.132$ &
		$10.512 \pm 0.070$ \\
		
		$\mu_{\alpha^\star}$ &
		$-12.461 \pm 0.049$ &
		$-12.444 \pm 0.049$ &
		$-12.186 \pm 0.060$ &
		$-12.200 \pm 0.026$ \\
		
		$\mu_\delta$ &
		$-0.979 \pm 0.044$ &
		$-0.988 \pm 0.044$ &
		$-1.027 \pm 0.046$ &
		$-1.454 \pm 0.020$ \\
		
		$\varpi$ &
		$1.480 \pm 0.192$ &
		$1.562 \pm 0.193$ &
		$1.661 \pm 0.190$ &
		$1.967 \pm 0.080$ \\
		
		\hline
		
		$D_\alpha$ &
		-- &
		$0.576 \pm 0.136$ &
		$0.198 \pm 0.144$ &
		$0.000 \pm 0.044$ \\
		
		$D_\delta$ &
		-- &
		$-0.034 \pm 0.088$ &
		$0.038 \pm 0.089$ &
		$0.000 \pm 0.032$ \\
		
		$\dot{D}_\alpha$ &
		-- &
		-- &
		$-0.001 \pm 0.000$ &
		$-0.131 \pm 0.022$ \\
		
		$\dot{D}_\delta$ &
		-- &
		-- &
		$0.000 \pm 0.000$ &
		$0.0129 \pm 0.0208$ \\
		
		$k$ &
		-- &
		-- &
		-- &
		$-3005 \pm 18450625^{(a)}$ \\
		
		$s$ &
		-- &
		-- &
		-- &
		$-48 \pm 18$ \\
		
		\hline
		
		$\ddot{D}_\alpha$ &
		-- &
		-- &
		-- &
		$-0.016 \pm 0.001$ \\
		
		$\ddot{D}_\delta$ &
		-- &
		-- &
		-- &
		$0.022 \pm 0.001$ \\
		
		$\Delta\mu_{\alpha^\star}$ &
		-- &
		-- &
		-- &
		$-0.758 \pm 0.002$ \\
		
		$\Delta\mu_\delta$ &
		-- &
		-- &
		-- &
		$-1.054 \pm 0.001$ \\
		
		\hline
	\end{tabular}
	\tablefoot{$^{(a)}$ The parameter $k$ (and hence also $s$) remains
		essentially unconstrained.}
\end{table*}

\begin{table*}
	\caption{Retrieved parameters for the additional $\pi^1$~Gru
		red-noise retrievals in which the convection-induced
		photocentre variability was injected directly from the
		three-dimensional radiation-hydrodynamic simulation
		\texttt{st29gm04n001} of \citet{Beguin_2024} and for which the companion flux $F_2$ was assumed unknown; see caption of Table~\ref{tab:input_retrieved_params}.}
	\label{tab:pi1gru_hydro_rednoise}
	\centering
	\begin{tabular}{lcc}
		\hline\hline
		Parameter & SS & VIMO \\
		\hline
		$\chi^2_{{\rm red,gen}}$ & 0.857 & 0.038 \\
		obj & $-1022$ & 31419 \\
		$\Delta \alpha^\star$ [mas]& $0.831 \pm 0.777$ & $0.369 \pm 0.753$ \\
		$\Delta \delta$ [mas]& $0.047 \pm 0.818$ & $0.438 \pm 0.770$ \\
		$\mu_{\alpha^\star}$ [mas a$^{-1}$] & $34.340 \pm 0.254$ & $30.964 \pm 0.718$ \\
		$\mu_\delta$  [mas a$^{-1}$] & $-20.310 \pm 0.228$ & $-18.153 \pm 0.451$ \\
		$\varpi$ [mas] & $5.819 \pm 0.225$ & $5.677 \pm 0.133$ \\
		$P$ [d] & -- & $4558 \pm 349$ \\
		$T_0$ [yr] & -- & $2032.48 \pm 0.90^{(a)}$ \\
		$\overline{\alpha}$ [mas] & -- & $21.38 \pm 2.33$ \\
		$e$ & -- & $0.050 \pm 0.055$ \\
		$\omega$ [$^\circ$] & -- & $248.9 \pm 24.8^{(a)}$ \\
		$i$ [$^\circ$] & -- & $22.4 \pm 7.3$ \\
		$\Omega$ [$^\circ$] & -- & $143.7 \pm 20.4$ \\
		$A_f$ & -- & $0.239 \pm 0.874$ \\
		$\sigma_{\rm jit}$ [mas]& $0.00001$ & $0.00001$ \\
		$\sigma_{{\rm red},\alpha^\star}$ [mas] & $2.00^{(b)}$ & $0.596$ \\
		$\tau_{{\rm red},\alpha^\star}$ [d] & $300^{(b)}$ & $84$ \\
		$\sigma_{{\rm red},\delta}$ [mas]& $2.00^{(b)}$ & $1.020$ \\
		$\tau_{{\rm red},\delta}$ [d] & $300^{(b)}$ & $93$ \\
		\hline
	\end{tabular}
	\tablefoot{
		(a) Because the simulated orbit is nearly circular
		($e \simeq 0$), the argument of periastron $\omega$ and
		time of periastron passage $T_0$ remain intrinsically
		ill constrained. 
		(b) Convergence toward imposed parameter boundary.
	}
\end{table*}

\begin{table*}[htp]
	\centering
	\caption{Retrieved single-star astrometric parameters for the simulated
		$\pi^1$~Gru Gaia DR2, DR3, and DR4 observing baselines.}
	\label{tab:pi1gru_dr2dr3dr4}
	\begin{tabular}{lccc}
		\hline\hline
		Parameter & DR2 & DR3 & DR4 \\
		\hline
		$\chi^2_{\rm red}$ & 3 & 7 & 93 \\
		\hline
		$\Delta\alpha^\star$ [mas]
		& $-32.027 \pm 0.070$
		& $-17.040 \pm 0.068$
		& $-9.604 \pm 0.235$ \\
		
		$\Delta\delta$ [mas]
		& $0.089 \pm 0.130$
		& $0.007 \pm 0.081$
		& $10.527 \pm 0.197$ \\
		
		$\mu_{\alpha^\star}$ [mas\,a$^{-1}$]
		& $29.235 \pm 0.109$
		& $31.047 \pm 0.079$
		& $34.925 \pm 0.130$ \\
		
		$\mu_{\delta}$ [mas\,a$^{-1}$]
		& $-9.098 \pm 0.157$
		& $-9.315 \pm 0.102$
		& $-12.640 \pm 0.147$ \\
		
		$\varpi$ [mas]
		& $5.327 \pm 0.102$
		& $5.526 \pm 0.107$
		& $5.041 \pm 0.284$ \\
		\hline
	\end{tabular}
\tablefoot{The fits were performed using a classical single-star astrometric
model without including red noise in the retrieval.
The quoted uncertainties were rescaled following the prescription of
El-Badry et al.\ (2024) assuming independent white-noise measurements.
The corresponding full VIMO retrieval results for the simulated
Gaia DR5 epoch astrometry are presented in
Table~\ref{tab:input_retrieved_params}.}
\end{table*}

\begin{table*}[htp]
	\centering
	\caption{Similar to Table~\ref{tab:pi1gru_dr2dr3dr4}, but for V~Hya}
	\label{tab:vhya_dr2dr3dr4}
	\begin{tabular}{lccc}
		\hline\hline
		Parameter & DR2 & DR3 & DR4 \\
		\hline
		$\chi^2_{\rm red}$ & 12 & 36 & 101 \\
		\hline
		
		$\Delta\alpha^\star$ [mas]
		& $1.613 \pm 0.088$
		& $-8.111 \pm 0.076$
		& $-8.256 \pm 0.088$ \\
		
		$\Delta\delta$ [mas]
		& $-6.528 \pm 0.055$
		& $-3.582 \pm 0.074$
		& $2.402 \pm 0.072$ \\
		
		$\mu_{\alpha^\star}$ [mas\,a$^{-1}$]
		& $-17.695 \pm 0.130$
		& $-16.190 \pm 0.111$
		& $-14.428 \pm 0.052$ \\
		
		$\mu_{\delta}$ [mas\,a$^{-1}$]
		& $2.144 \pm 0.080$
		& $1.996 \pm 0.073$
		& $1.161 \pm 0.038$ \\
		
		$\varpi$ [mas]
		& $1.073 \pm 0.076$
		& $1.864 \pm 0.098$
		& $2.126 \pm 0.118$ \\
		\hline
	\end{tabular}
\end{table*}

\begin{table*}[htp]
	\centering
	\caption{Similar to Table~\ref{tab:input_retrieved_params}, but now for Gaia DR4 epoch astrometry.}
	\label{tab:pi1gru_vhya_dr4_rednoise}
	\centering
	\begin{tabular}{l|cc|cc}
		\hline\hline
		Parameter 
		& \multicolumn{2}{c|}{$\pi^1$~Gru}
		& \multicolumn{2}{c}{V~Hya} \\
		\cline{2-3}\cline{4-5}
		& SS & VIMO ($F_2$ unknown)
		& SS & VIMO ($F_2$ unknown) \\
		\hline
		
		$\chi^2_{\rm red,gen}$ 
		& 0.194 & 0.010
		& 0.065 & 0.025 \\
		
		obj
		& $-1086$ & $-1373$ 
		& $-3308$ & $-3321$ \\
		\hline
		
		$\Delta \alpha^\star$ [mas] &
		$-7.880 \pm 1.180$ &
		$-4.826 \pm 0.572$ &
		$-8.205 \pm 0.836$ &
		$-6.949 \pm 0.910$ \\
		
		$\Delta \delta$ [mas]&
		$8.110 \pm 1.660$ &
		$4.730 \pm 1.120$ &
		$2.037 \pm 0.676$ &
		$-0.820 \pm 1.250$ \\
		
		$\mu_{\alpha^\star}$ [mas a$^{-1}$]&
		$36.161 \pm 0.509$ &
		$35.829 \pm 0.180$ &
		$-14.270 \pm 0.405$ &
		$-14.381 \pm 0.331$ \\
		
		$\mu_\delta$  [mas a$^{-1}$]&
		$-13.111 \pm 0.573$ &
		$-14.357 \pm 0.721$ &
		$1.152 \pm 0.346$ &
		$1.345 \pm 0.198$ \\
		
		$\varpi$ [mas]&
		$5.758 \pm 0.316$ &
		$5.851 \pm 0.084$ &
		$2.534 \pm 0.279$ &
		$2.405 \pm 0.231$ \\
		
		\hline
		
		$P$ [d] &
		-- &
		$2082 \pm 133$ &
		-- &
		$584.170 \pm 9.800$ \\
		
		$T_0$ [yr] &
		-- &
		$2031.705 \pm 0.299^{(b)}$ &
		-- &
		$2021.405 \pm 0.018^{(b)}$ \\
		
		$\overline{\alpha}$ [mas]&
		-- &
		$8.066 \pm 0.833$ &
		-- &
		$5.200 \pm 14.600$ \\
		
		$e$ &
		-- &
		$0.484 \pm 0.052$ &
		-- &
		$0.950 \pm 0.315$ \\
		
		$\omega$ [$^\circ$] &
		-- &
		$176.639 \pm 4.120$ &
		-- &
		$288.715 \pm 69.300$ \\
		
		$i$ [$^\circ$] &
		-- &
		$56.550 \pm 4.800$ &
		-- &
		$109.900 \pm 52.600$ \\
		
		$\Omega$ [$^\circ$] &
		-- &
		$316.149 \pm 5.860$ &
		-- &
		$111.552 \pm 36.000$ \\
		
		$A_f$ &
		-- &
		$1.067 \pm 1.230$ &
		-- &
		$0.193 \pm 0.119$ \\
		
		\hline
		
		$\sigma_{\rm jit}$ [mas]&
		0.00001 &
		0.00001 &
		0.00001 &
		0.00001 \\
		
		$\sigma_{\rm red,\alpha^\star}$ [mas]&
		$2.000^{(a)}$ &
		$0.29$ &
		$1.59$ &
		$1.26$ \\
		
		$\tau_{\rm red,\alpha^\star}$ [d] &
		$300^{(a)}$ &
		$77$ &
		$300^{(a)}$ &
		$300^{(a)}$ \\
		
		$\sigma_{\rm red,\delta}$ [mas]&
		$2.000^{(a)}$ &
		$0.169$ &
		$1.497$ &
		$0.842$ \\
		
		$\tau_{\rm red,\delta}$ [d] &
		$300^{(a)}$ &
		$32$ &
		$218$ &
		$179$ \\
		
		\hline
	\end{tabular}
	\tablefoot{
		The goodness-of-fit values for the red-noise models are computed using
		the full covariance matrix and therefore correspond to generalized,
		rather than classical white-noise, reduced chi-square statistics. \\
		$^{(a)}$ Convergence toward imposed parameter boundary.
		$^{(b)}$ The quoted $T_0$ values are expressed as the future equivalent
		epochs listed by the orbital fit.
	}
\end{table*}

\end{appendix}
\end{document}